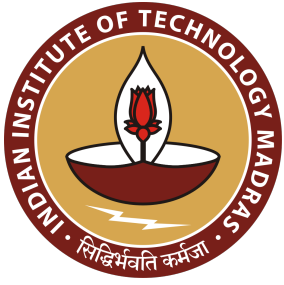

DEPARTMENT OF AEROSPACE ENGINEERING
INDIAN INSTITUTE OF TECHNOLOGY MADRAS
CHENNAI – 600036

# Dynamics of Coupled Metamaterials: Acoustic Black Hole, Local Resonator & Multistable Oscillator

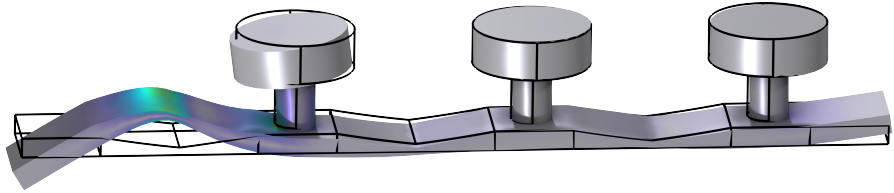

*A Thesis*

*Submitted by*

**ARGHYA MONDAL**

*For the award of the degree*

*Of*

**MASTER OF SCIENCE**

January 2024

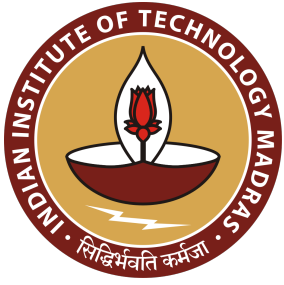

DEPARTMENT OF AEROSPACE ENGINEERING
INDIAN INSTITUTE OF TECHNOLOGY MADRAS
CHENNAI – 600036

# Dynamics of Coupled Metamaterials: Acoustic Black Hole, Local Resonator & Multistable Oscillator

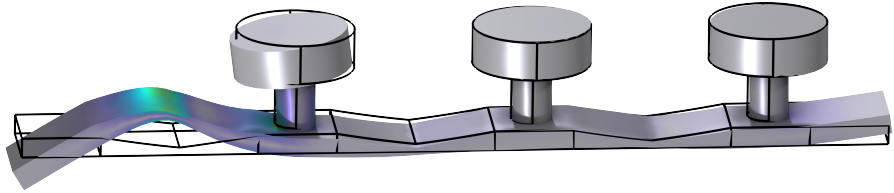

*A Thesis*

*Submitted by*

**ARGHYA MONDAL**

*For the award of the degree*

*Of*

**MASTER OF SCIENCE**

January 2024

*If you want to find the secrets of the universe, think in terms of energy, frequency and vibration.*

**– Nikola Tesla**

*Dedicated to the people who always motivated me to pursue my true passion and helped me to stay strong at tough times.*

# THESIS CERTIFICATE

This is to undertake that the Thesis titled **DYNAMICS OF COUPLED METAMATERIALS: ACOUSTIC BLACK HOLE, LOCAL RESONATOR & MULTISTABLE OSCILLATOR**, submitted by me to the Indian Institute of Technology Madras, for the award of **Master of Science**, is a bona fide record of the research work done by me under the supervision of **Dr. Senthil Murugan**. The contents of this Thesis, in full or in parts, have not been submitted to any other Institute or University for the award of any degree or diploma.

**Chennai 600036**  **Arghya Mondal**

**Date: January 2024**

**Dr. Senthil Murugan**
Research Advisor
Assistant Professor
Department of Aerospace Engineering
Indian Institute of Technology Madras

# LIST OF PUBLICATIONS

### I. REFEREED JOURNALS BASED ON THESIS

Mondal, Arghya, and Senthil Murugan. "Flexural wave propagation characteristics of metabeam with simultaneous acoustic black hole and local resonator." European Journal of Mechanics-A/Solids (2023): 105217. *DOI:* https://doi.org/10.1016/j.euromechsol.2023.105217

### II. PUBLICATIONS IN CONFERENCE PROCEEDINGS

Mondal, Arghya, Sayan Dutta, and Senthil Murugan. "Coupled flexural and torsional vibration attenuation with locally resonant metamaterials." Materials Today: Proceedings (2023). *DOI:* https://doi.org/10.1016/j.matpr.2023.01.111

### III. PRESENTATIONS IN CONFERENCES

A. Mondal, S. Dutta & S. Murugan, Coupled flexural and torsional vibration attenuation with LR metamaterials, 13$^{th}$ *International Symposium on Plasticity and Impact Mechanics (August 2022)*, Indian Institute of Technology Madras.

A. Mondal & S. Murugan, Attenuation bandwidth enhancement in meta-structures with nonlinear multistable local resonator, 18$^{th}$ *International Conference on Vibration Engineering & Technology Of Machinery (December 2023)*, Indian Institute of Technology Roorkee.

# ACKNOWLEDGEMENTS

First and foremost, I would like to express my deepest gratitude to Professor Senthil Murugan, my thesis advisor, for providing me with the opportunity to pursue this research and for his continuous guidance over the past two years at the Indian Institute of Technology Madras.

I am also thankful to Professor David Kumar and Professor Ganesh Tamadapu for agreeing to serve on my thesis committee with their support and insightful comments. Special thanks to the Head of the Aerospace Engineering Department, Professor H S N Murthy, for granting access to research facilities.

My sincere appreciation goes to my colleagues at the Aero-Electro-Mechanics and Systems laboratory for sharing their knowledge and passion for structural dynamics and, above all, for creating an invaluable atmosphere in the lab. I extend my thanks to my family and friends. I am grateful to my parents for their understanding of my distance from them and for their unwavering support whenever needed. I also appreciate my friends for the meals and parties we had together. Many of us will attend commencement together, creating yet another wonderful memory.

# ABSTRACT

Vibration attenuation has played a crucial role in engineering structure, wherein the innovative structure and metamaterials have found escalated usage. These structures can be cleverly built to be lightweight and have negative mass properties by which waves can be attenuated at specific frequency bands.

The first part studies the flexural wave propagation in a metamaterial beam with the coupled acoustic black hole and local resonators (ABHLR). The main objectives are to derive the dynamic stiffness matrix of the coupled ABHLR unit cell and analyse its wave propagation characteristics. Initially, a closed-form solution of Euler-Bernoulli's beam with an indented acoustic black hole is derived. From this, an exact dynamic stiffness matrix, incorporating local resonator effects, is derived using the spectral element method. Dispersion diagrams are generated by combining Bloch's theorem with the ABHLR transfer matrix. The coupled effects of acoustic black hole (ABH) and local resonator (LR) on the bandgap characteristics are investigated. The coupled ABHLR metamaterial shows either a positive or negative effect on bandgap formations, depending on the parametric values of the unit cell. As a positive effect, for some parametric values, the coupled ABHLR cell simultaneously produces bandgaps at lower frequencies (LR-based) and higher frequency ranges (ABH-based). Similarly, for other parametric values, as a negative effect, the bandgaps of the ABHLR cell get merged, but the pass bands appear on the overlapping region. This pass-band appears due to the constructive interference of ABH and LR on each other. However, the attenuation bandgap with a higher attenuation rate is obtained on the overlapping region due to the destructive interference of LR and ABH effect. The global stiffness matrix of a finite beam with 30 unit cells is developed to analyse the frequency domain response, and the results are validated with band gap analysis of the unit cell. The frequency domain response is also validated with finite element simulations using commercial software. Finally, the ABHLR beam is fabricated employing 3D printing technology, and the frequency response is analysed experimentally. Results show that ABH and LR parameters have to be optimally designed to utilise the attenuation effects and avoid constructive interference due to coupling effects. The coupling of metamaterials can aid in vibration suppression and multiple-frequency filtering in mechanical and aerospace applications.

The second part explores the impact of multistable nonlinear oscillators in a 1D metamaterial chain to enhance the attenuation bandwidth of the meta-structure. Bloch's theorem cannot be directly implemented in nonlinear metamaterials due to their amplitude dependency. A time-domain study is conducted using the Range-Kutta method to solve finite nonlinear metamaterial chains. Furthermore, Frequency Response Function (FRF) studies (up and down frequency sweeps) are analyzed using root mean square values of steady-state displacements at each frequency. The results reveal a correlation between the potential well of multistable oscillators and the attenuation bandwidth of metamaterial structures. Three types of oscillators are employed based on the number of stable points: 1) Monostable, 2) Bistable, and 3) Tristable. Multistable oscillators reveal intrawell and interwell oscillations at low and high excitation levels, respectively. It is observed that intrawell oscillations create a single-attenuation band region, depending on the initial stability point. In contrast, interwell oscillations generate multiple attenuation band regions, which depend on the number of nonidentical potential wells of the multistable oscillator or resonator. These multistable oscillators can also be utilized for multiple-frequency vibration filtering.

# CONTENTS

# LIST OF TABLES

# LIST OF FIGURES

# ABBREVIATIONS

**ABH**      Acoustic Black Hole.

**ABHLR**   Coupled Acoustic Black Hole and Local Resonator.

**BS**       Bragg's Scattering.

**DS**       Dynamic Stiffness.

**FEM**     Finite Element Method.

**FRF**     Frequency Response Function.

**LR**       Local Resonator.

**RMS**     Root Mean Square.

**SEM**     Spectral Element Matrix.

**TM**      Transfer Matrix.

# CHAPTER 1

# INTRODUCTION

## 1.1 LITERATURE REVIEW

Metamaterials or periodic structures have received increased interest from researchers due to their implications in phononics, composites, acoustics, and mechanics. The historical roots of studies on wave propagation via periodic structures have been started from Rayleigh's investigation into continuous periodic structures and Newton's work on clarifying sound propagation in air. In recent decades, structural experts have become more interested in artificially constructed periodic structures due to their innovative capabilities of vibration reduction and particular frequency filtering.

### 1.1.1 Mechanical Metamaterial

Mechanical metamaterials are a class of materials engineered to exhibit unique and exceptional mechanical properties that are not found in conventional materials. The term "metamaterial" is derived from "meta", meaning "go beyond", indicating that these materials go beyond the properties of conventional materials. Metamaterials are capable of exhibiting extreme mechanical properties, including negative Poisson's ratio, negative effective mass density, and negative effective stiffness. Periodic metamaterials that are designed to control the propagation of elastic or sound waves are also known as phononic crystals. Periodic arrangements of phononic crystals exhibit forbidden bands or bandgaps due to Bragg's scattering of elastic waves with wavelength comparable to the dimension of the period of the crystal. Later, a new class of metamaterial named locally resonant structure that generated a bandgap independent of the lattice size constraint was presented theoretically and experimentally by Liu *et al.* (2000). This type of metamaterial brings flexibility in designing the meta-structure, especially for acoustic wave filtering, as the sonic waves are an order of magnitude larger than the lattice constant of the structure. These types of metamaterial with local resonators are called acoustic metamaterial. Phononic or acoustic metamaterials can generate a propagation band (waves can propagate through these frequencies region) and attenuation band (waves can not propagate through these frequencies region) zones due to local resonance, Bragg's scattering or their coupling in general. Local resonance develops due to out-of-phase resonance, while Bragg's scattering results from destructive interference phenomena, as discussed by Liu and Hussein (2012).

### 1.1.2 Acoustic Black Hole

The acoustic black hole is a passive vibration control technique designed to trap and absorb acoustic waves, just like a black hole in astrophysics traps light due to its strong gravitational pull. ABH effect can be achieved by embedding a geometrical in-homogeneity in a structure, typically a beam or plate, according to the power-law profile discussed by Tang (2017). Mironov (1988) first invented a power-law profiled thickness varying beam having a flexural wave-trapping property, zero phases and group velocity at the idealized ABH termination point. Practically, a truncated tip in ABH thickness results in inevitable wave reflection. By combining the ABH effect with the embedding of an absorbent layer, it is possible to minimise wave reflection caused by truncated edges, as proved experimentally by Krylov (2004); Krylov and Winward (2007). Researchers like Pelat *et al.* (2020); Karlos (2020) have investigated various aspects of 1D and 2D ABH structures and analysed the effectiveness of single ABH theoretically and experimentally. Several analytical and semi-analytical methods have been proposed for 1D and 2D configurations, revealing insights into dominant wave propagation phenomena by Lee and Jeon (2019); Karlos *et al.* (2019); Deng *et al.* (2023); Deng and Gao (2022); Deng *et al.* (2022); Ma and Deng (2022). With applications like vibro-impact systems by Li *et al.* (2019), cochlear systems, and acoustic ducts with a variable cross-sectional area by Guasch *et al.* (2017, 2020); Deng *et al.* (2021), energy harvesting by Zhao *et al.* (2014*a*), acoustic black holes have paved the way for the synthesis of novel metamaterials used in waveguides and wave tailoring. Flexural wave propagation in beam structures with embedded periodic ABH has been initially investigated by Tang and Cheng (2017*a*). Embedding multiple ABH can improve wave suppression to some extent in mid-high frequency regions.

### 1.1.3 Multistable Oscillator

Nonlinear metamaterials have emerged as a promising area of research in structural dynamics, particularly for the suppression of structural vibrations. Unlike traditional linear resonator metamaterials, multistable oscillator units can exhibit intriguing and controllable nonlinear responses, enabling a wide range of attenuation band regions. Nonlinear resonance can increase bandgaps by using properties such as sub and super-harmonic resonances, period multiplication, and chaotic response. Multistability is a nonlinear resonance phenomenon that is influenced by disturbances, initial conditions, system characteristics, and other factors, as discussed by Fang *et al.* (2022). Ferromagnetic beams and beams with tip magnets in the presence of a magnetic field can generate monostable, bistable, or multistable oscillations depending on the position of the magnets, as investigated by Harne and Wang (2013); Zhou *et al.* (2014); Kim and Seok (2014); Krylov and Winward (2007). Multistable systems can also be found in adaptive composite structures embedding piezoelectric materials, as explored by Lee *et al.* (2017), or shape memory alloys, as studied by Kim *et al.* (2014). Other instances include pre-stressed composite structures with an initial curvature, as demonstrated by Coburn *et al.* (2013), and origami structures, as discussed by Hanna *et al.* (2014).

According to Banerjee *et al.* (2016), the effects of cubic duffing type non-linearity on 1D elastodynamic metamaterials revealed that a bistable nonlinear system has a broader attenuation bandwidth than a monostable system. Xia *et al.* (2019) observed that a base-excited cantilever beam with magneto-elastic bistable unit cells has a much wider attenuation bandwidth due to nonlinear interwell oscillations than the linear locally resonant bandgap.

## 1.2 MOTIVATION

Vibration isolation has played an essential role in various domains like structural engineering, aerospace and automotive industry. Metamaterials, engineered structures with unique properties, have garnered significant attention in the realm of vibration attenuation. In the aerospace industry, metamaterials can be designed to attenuate vibrations in aircraft components, thereby ensuring improved operational performance and heightened passenger comfort. Researchers have developed periodic phononic metamaterials incorporating local resonators to enhance low-frequency vibration attenuation. Consequently, simultaneous attenuation of multiple frequencies in specific structures has become essential with the advancement of technology.

To accomplish this objective, various mechanisms for bandgap formation can be coupled to generate multiple attenuation bands. This coupling effect can manifest distinct physical behaviours as a result of the merging of band structures. The coupled effects of Acoustic Black Holes and Local Resonators in metamaterial beams have been thoroughly studied to gain better insights into the coupling behaviour. On the other side, multistable oscillators in metamaterial structures have captured the attention of researchers due to their multistable behaviour. In this thesis, it has been explored that these multistable oscillators can open up multiple attenuation bands depending on their stability points.

## 1.3 OBJECTIVE 1: COUPLED ACOUSTIC BLACK HOLE & LOCAL RESONATOR

Flexural vibrations or waves are a type of mechanical wave that propagates in structures when subjected to a bending moment or load. Mechanical metamaterials can be implemented inside the structure to attenuate this type of vibration. Typically, local resonators are installed periodically to enhance low-frequency attenuation. To reduce the additional mass caused by local resonators, acoustic black holes can be embedded periodically throughout the structure. Acoustic black holes also enhance the attenuation by creating periodic inhomogeneity. These coupled ABHLRs (Acoustic black hole and local resonator) make structures lightweight and suitable for attenuating multiple frequencies.

In this thesis, flexural vibration characteristics of metamaterial beam with coupled acoustic black hole and local resonator have been studied. The coupled ABHLR shows a specific wave propagation effect due to the overlapping of two different band formation mechanisms. The local resonance effect interferes with the wave-trapping property of the acoustic black hole, yielding specific wave propagation characteristics. The coupled ABHLR can have both positive and negative effects on the band formation. The characteristics of ABHLR beam have been analysed rigorously so that coupling can be optimally designed to maximize the attenuation effects.

## 1.4 OBJECTIVE 2: MULTISTABLE OSCILLATOR

Throughout the last decade, researchers have been using metamaterial chains for longitudinal vibration attenuation. Most of them have focused on metamaterial chains with linear local resonator systems. Recently, nonlinear multistable oscillators have claimed attention because of their sub and super-harmonic resonances, period multiplication and chaotic response. Multistability is another nonlinear effect which is renowned for having more than one stable point.

In this thesis, a 1D metamaterial chain with nonlinear multistable oscillators has been studied for longitudinal vibration attenuation. There is a relation between the attenuation band and the potential energy well of the multistable oscillators. Multistable oscillators can open up multiple attenuation band zones depending on the stability points. To understand the better insights of nonlinear regimes, a metamaterial chain with multistable oscillators has been analysed numerically.

**1.5 OUTLINE OF THESIS**

This thesis commences with a literature survey, presenting a substantial amount of information on developments in this field in the first chapter. It outlines the major theories and the nature of the development. The discussion identifies gaps in the literature, concluding with a brief overview of the analysis. Chapter 2 explores metamaterial theory, encompassing the acoustic black hole effect. Chapter 3 clearly explains the wave propagation analysis method utilized in this thesis. These elements contribute to a strong understanding for readers before proceeding to the next chapter.

Chapter 4 introduces the spectral element method formulation of the coupled ABHLR cell for flexural vibration attenuation. This method is further validated through finite element simulations in the next chapter. Chapter 5 focuses on the results section, and conclusions are drawn from the coupled ABHLR cell. The experimental investigation of this model is detailed in Chapter 6. Chapter 7 also explains a basic study of a metamaterial chain with multistable oscillators for longitudinal vibration attenuation. This study can be explored rigorously in future. Chapter 8 summarizes the overall conclusions and outlines future directions for this research.

# CHAPTER 2

# METAMATERIAL THEORY

## 2.1 WAVE PROPAGATION & CONTROL

Acoustic metamaterials are materials designed to regulate and modify elastic wave propagation in novel ways inside solids. In order to understand how vibrations are attenuated via solids, we must first introduce and familiarise ourselves with the characteristics of the wave equation. The fundamental components for controlling wave propagation through periodic metamaterial include: Bragg's Scattering and Local Resonator. In this section, we briefly explore these two principles in the context of contemporary metamaterial design.

### 2.1.1 Bragg's Scattering

Waves can be characterized as the travelling modulation of a certain property *f(x,t)*, such as

$$f(x,t) = \cos(\omega t - kx + \varphi), \tag{2.1}$$

where $\omega = 2\pi/T$ represents the angular frequency with period $T$, $k = 2\pi/\lambda$ represents the wave number with wave-length $\lambda$, and $\varphi$ represents a phase.

Waves exhibiting a phase difference $\Delta\varphi$ can undergo interference, resulting in constructive or destructive effects depending on this phase discrepancy. Destructive interference occurs when the phase difference between two waves is $\pi$, while constructive interference occurs when the phase difference is zero. On the other hand, When a wave impinges upon a scatterer, the scatterer becomes excited, and it subsequently radiates waves. Fig. 2.1 illustrates two identical scatterers separated by a distance $\Delta x$, and these emit waves with a phase difference. When the phase difference between two scattered waves, expressed as $k\Delta x$, equals $\pi$, the scattered waves undergo destructive interference.

For the above condition, the separation distance $\Delta x$ becomes $\frac{\lambda}{2}$, which shows a destructive interference effect on the wave propagation.

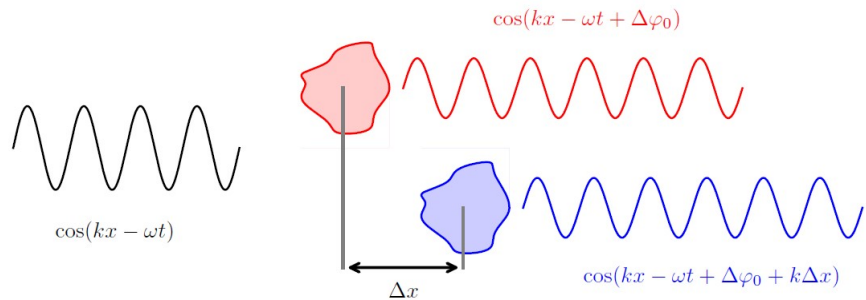

Figure 2.1: Scattered wave from two identical scatterers separated by $\Delta x$ distance, as shown by Huber (2018)

This phenomenon, known as "Bragg's scattering", is observed when dealing with periodic arrays of scatterers. Phononic crystals are fundamentally associated with Bragg's scattering. It can be easily visualized that at frequencies where the Bragg condition, $\Delta L = \frac{\lambda}{2}$, is satisfied, periodic arrays with a characteristic length scale $\Delta L$ are showing non-propagating wave-band region.

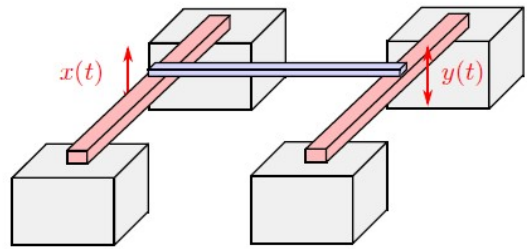

Figure 2.2: Two coupled oscillators, as shown by Huber (2018)

### 2.1.2 Local Resonator

The local resonator is another building element designed to produce a similar effect while avoiding structural size limits. To comprehend the idea of local resonances, we must first understand what happens as two oscillators are coupled. Fig. 2.2 shows two coupled oscillator systems. The displacements of the oscillators *x(t)* and *y(t)* are defined

as

$$\ddot{x}(t) = -\omega_0^2 x(t) + \gamma^2 y(t)$$
$$\ddot{y}(t) = -\omega_0^2 y(t) + \gamma^2 x(t)$$
(2.2)

Where $\omega_0$ is the natural frequency of two oscillators, and $\gamma$ is the coupling constant. The eigenfrequencies of the above system can be derived as

$$\omega_\pm = \sqrt{\omega_0^2 \pm \gamma^2} \tag{2.3}$$

If two coupled oscillators have different natural frequencies, eigenfrequencies can be calculated as

$$\omega_\pm = \sqrt{\left(\frac{\omega_x^2 + \omega_y^2}{2}\right)^2 \pm \left[\left(\frac{\omega_x^2 - \omega_y^2}{2}\right)^2 + \gamma^2\right]^2} \tag{2.4}$$

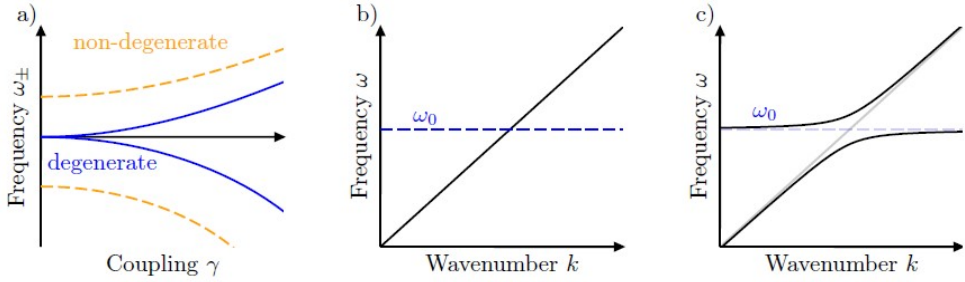

Figure 2.3: a) Effect of the coupling parameter $\gamma$ on the eigenfrequency. b) Wave dispersion (black) and the position of a local resonance (blue). c) Coupling the local resonance leads to the opening of a band gap, as discussed by Huber (2018).

Fig. 2.3a illustrates how the coupling constant $\gamma$ affects the eigenfrequency of two oscillators. The coupling effect between two oscillators is strongest when they are degenerate, suggesting that they have the same frequency. Furthermore, the primary impact of $\gamma$ is to cause frequency splitting in the two-oscillator system. Fig. 2.3b and Fig. 2.3c illustrate how local resonance affects wave dispersion. The coupling effect splits the wave of the degenerate system and the local oscillator, with the highest influence

occurring at the junction of their frequencies. A frequency window opens where waves don't propagate, similar to the Bragg scattering phenomenon. Notably, in this situation, the non-propagating wave frequency depends on the frequency of the local oscillator instead of the periodic array spacing.

## 2.2 ACOUSTIC BLACK HOLE

The Acoustic Black Hole (ABH) creates zero reflection of propagating waves by gradually reducing both phase and group velocity to zero at the zero-thickness position. However, in practical scenarios, the thickness does not reach zero, allowing some waves to propagate through the remaining thickness. Fig. 2.4 illustrates an acoustic black hole-embedded 2D plate with a nonzero residual thickness. The variable thickness profile follows a power-law relation expressed as $h(x) = \varepsilon x^m + h_1$, where $\varepsilon$ is a constant, and $m$ represents the order of power.

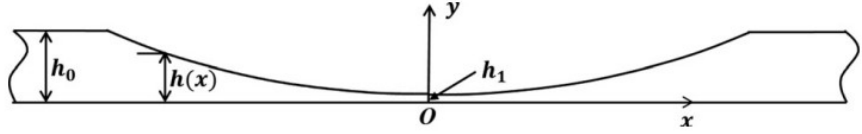

Figure 2.4: ABH plate with nonzero residual thickness $h_1$, as illustrated by Zhao (2016).

The flexural wave propagation through the ABH plate (Krylov (2004)) in $x$-direction can be expressed as

$$A(x) = B(x)\exp(i\theta(x)) \qquad (2.5)$$

where $B(x)$ represents variable amplitude, $i$ represents the imaginary unit, and $\theta(x) = \int_0^x k(x)dx$ depicts the total accumulated phase. The wave number $k(x)$ of the propagating flexural wave can be represented as

$$k(x) = \sqrt[4]{12}\sqrt{\frac{k_l}{\varepsilon x^m + h_1}} \qquad (2.6)$$

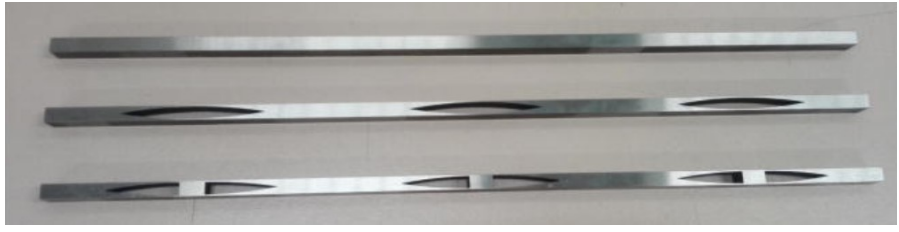

Figure 2.5: Beam with multiple ABH with or without reinforcing stud, as shown by Tang and Cheng (2017*b*).

where $k_l^2 = \rho \left(1 - v^2\right) \omega^2 / E$ signifies the longitudinal wave number for a plate with constant thickness. When waves propagate through the plate, the wave number ($k(x)$) tends towards infinity as the residual thickness approaches zero at the middle part of the plate. Consequently, the power-law profiled thickness induces a sweeping of wave numbers, progressively reducing the wavelength of travelling waves.

The group velocity $c_g$ of the propagating wave can be derived in such a structure, as discussed by Zhao *et al.* (2014*b*). These can be written as

$$c_g = \sqrt[4]{\frac{4E}{3\rho \left(1 - v^2\right)}} \sqrt{\omega \cdot (\varepsilon x^m + h_1)} \qquad (2.7)$$

where $\omega$ represents the angular frequency of the flexural wave, $E$ denotes Young's modulus, and $\rho$ represents the density. When the waves are in the middle of the ABH and the remaining thickness $h_1$ approaches zero, the group velocity ($c_g$) tends to zero. This means that the middle area of the ABH becomes a singularity point where particle displacement approaches infinity. Meanwhile, vibrations outside the ABH will diminish due to energy equilibrium. This suggests that the tapered ABH can serve as a useful passive vibration control device.

In practice, it is impossible to produce power-law profiles with zero residual thickness. So, a part of the propagating waves can pass through the remaining thickness, resulting in significantly decreased vibration control capabilities. Later, periodic lattices with embedded ABH were investigated for controlling the transient responses. In general, the periodic arrangement of ABH can be considered as a metamaterial, with the ABH acting as the fundamental unit cell. Periodic ABH metamaterial can significantly reduce the propagating wave over a single ABH absorber.

# CHAPTER 3

# WAVE PROPAGATION ANALYSIS METHOD

Throughout this thesis, we employed the spectral element approach to analyse wave propagation. In this chapter, we covered how to construct the dynamic stiffness matrix and transfer matrix for a basic Euler-Bernoulli beam. The spectral element matrix methods are explained in detail by the book Lee (2009). Here, it is discussed in brief to get a fundamental understanding. Furthermore, these approaches are used for wave propagation analysis of a metamaterial beam containing coupled acoustic black holes and local resonators in the next Chapter.

## 3.1 SPECTRAL ELEMENT METHOD

The spectral element approach combines the finite element, dynamic stiffness, and spectral analysis methods. The spectral element approach is applied using the following processes:

- The time-domain governing partial differential equation is converted into a frequency-domain ordinary differential equation.

- The ordinary differential equation in the frequency domain is solved accurately, and the exact solution is used to generate frequency-dependent dynamic form functions.

- The spectral element matrix is constructed from the dynamic shape functions in the same way as the finite element approach.

The spectral element matrix can be created using three approaches. The fundamental characteristics of each approach have been outlined below:

1. **Force displacement relation method:** This approach yields an exact solution to the governing equation of motion. The spectral element matrix is then created by establishing a direct link between nodal forces and nodal displacement.

2. **Variational method:** The variational technique is typically employed for

multidimensional problems.

3. **State-vector equation method:** The preceding two techniques are only applicable when exact wave solutions to the frequency domain governing equations are known in closed forms. This approach does not require exact solutions.

### 3.1.1 Dynamic Stiffness Matrix

In this thesis, we used the force-displacement relation approach to calculate the spectral element matrix or dynamic stiffness matrix for the selected model. Here, we generated the spectral element matrix for the flexural vibration of an Euler-Bernoulli beam, as illustrated by Lee (2009), to build a fundamental understanding. The free flexural vibration of an Euler-Bernoulli beam can be described by:

$$EIw'''' + \rho A \ddot{w} = 0 \tag{3.1}$$

where $A$ is the cross-sectional area, $I$ is the area moment of inertia around the neutral axis, $w(x,t)$ is the transverse displacement, $E$ is Young's modulus, and $\rho$ is the mass density. The bending moment and internal transverse shear force are provided by

$$M_t(x,t) = EIw''(x,t), \quad Q_t(x,t) = -EIw'''(x,t) \tag{3.2}$$

Assume the solution in the spectral form to be

$$w(x,t) = \frac{1}{N} \sum_{n=0}^{N-1} W_n(x;\omega_n) e^{i\omega_n t} \tag{3.3}$$

Once the solution mentioned above is substituted, the eigenvalue problem for a given discrete frequency can be expressed as

$$EIW'''' - \omega^2 \rho A W = 0 \tag{3.4}$$

Assuming the general solution as $W(x) = a e^{-ik(\omega)x}$ and substituting this solution yields a dispersion relation as

$$k^4 - k_F^4 = 0 \tag{3.5}$$

where $k_F = \sqrt{\omega}\left(\frac{\rho A}{EI}\right)^{1/4}$ is the wave number for the flexural wave mode. Both pure real and pure imaginary roots exist for this wave number. Thus, the general solution can be expressed as follows:

$$W(x;\omega) = a_1 e^{-ik_F x} + a_2 e^{-k_F x} + a_3 e^{+ik_F x} + a_4 e^{+k_F x} = \boldsymbol{e}(x;\omega)\boldsymbol{a} \tag{3.6}$$

where

$$\boldsymbol{e}(x;\omega) = \begin{bmatrix} e^{-ik_F x} & e^{-k_F x} & e^{+ik_F x} & e^{+k_F x} \end{bmatrix}$$

$$\boldsymbol{a} = \begin{Bmatrix} a_1 & a_2 & a_3 & a_4 \end{Bmatrix}^T$$

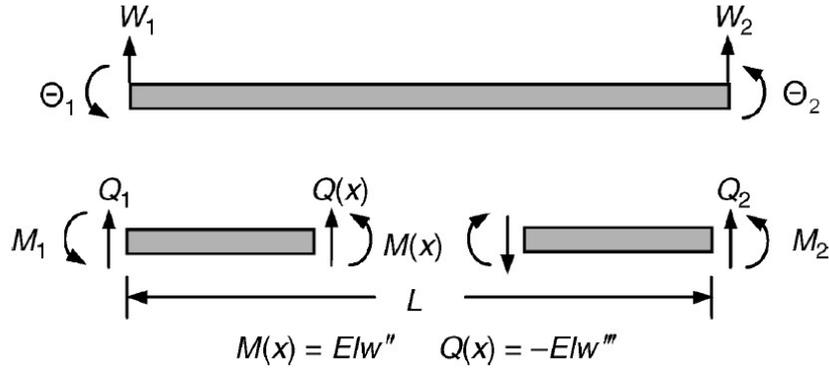

Figure 3.1: Sign convention for the Euler-Bernoulli beam element, as shown by Lee (2009)

The displacement field can be connected to the spectral nodal displacements and slopes of the finite beam element by

$$\boldsymbol{d} = \begin{Bmatrix} W_1 \\ \Theta_1 \\ W_2 \\ \Theta_2 \end{Bmatrix} = \begin{Bmatrix} W(0) \\ W'(0) \\ W(L) \\ W'(L) \end{Bmatrix}, \boldsymbol{d} = \begin{bmatrix} \boldsymbol{e}(0;\omega) \\ \boldsymbol{e}'(0;\omega) \\ \boldsymbol{e}(L;\omega) \\ \boldsymbol{e}'(L;\omega) \end{bmatrix} \boldsymbol{a} = \boldsymbol{H}_B(\omega)\boldsymbol{a} \tag{3.7}$$

where,

$$H_B(\omega) = \begin{bmatrix} 1 & 1 & 1 & 1 \\ -ik_F & -k_F & ik_F & k_F \\ e^{-ik_F L} & e^{-k_F L} & e^{+ik_F L} & e^{+k_F L} \\ -ik_F e^{-ik_F L} & -k_F e^{-k_F L} & ik_F e^{+ik_F L} & k_F e^{+k_F L} \end{bmatrix}$$

By eliminating the constant vector $a$, the displacement field can be expressed in terms of the nodal DOFs vector $d$. Consequently,

$$W(x) = N_B(x;\omega)d \qquad (3.8)$$

where,

$$N_B(x;\omega) = e(x;\omega)H_B^{-1}(\omega) = \begin{bmatrix} N_{B1} & N_{B2} & N_{B3} & N_{B4} \end{bmatrix}.$$

The force field vector can be connected to the appropriate nodal forces and moments, denoted as

$$f_c = \begin{Bmatrix} Q_1 \\ M_1 \\ Q_2 \\ M_2 \end{Bmatrix} = \begin{Bmatrix} -Q(0) \\ -M(0) \\ +Q(L) \\ +M(L) \end{Bmatrix} \qquad (3.9)$$

The force-displacement relation is $S_B(\omega)d = f_c(\omega)$, where $S_B(\omega)$ represents the spectral element matrix for the Euler-Bernoulli beam element. It can be stated as

$$S_B(\omega) = \frac{EI}{L^3} \begin{bmatrix} s_{B11} & s_{B12} & s_{B13} & s_{B14} \\ s_{B12} & s_{B22} & s_{B23} & s_{B24} \\ s_{B13} & s_{B23} & s_{B33} & s_{B34} \\ s_{B14} & s_{B24} & s_{B34} & s_{B44} \end{bmatrix} = S_B(\omega)^T \qquad (3.10)$$

where

$$s_{B11} = s_{B33} = \Delta_B \bar{L}^3(\cos \bar{L} \sinh \bar{L} + \sin \bar{L} \cosh \bar{L})$$

$$s_{B22} = s_{B44} = \Delta_B \bar{L}^3 k_F^{-2}(-\cos \bar{L} \sinh \bar{L} + \sin \bar{L} \cosh \bar{L})$$

$$s_{B12} = -s_{B34} = \Delta_B \bar{L}^3 k_F^{-1} \sin \bar{L} \sinh \bar{L}$$

$$s_{B13} = -\Delta_B \bar{L}^3 (\sin \bar{L} + \sinh \bar{L})$$

$$s_{B14} = -s_{B23} = \Delta_B \bar{L}^3 k_F^{-1}(-\cos \bar{L} + \cosh \bar{L})$$

$$s_{B24} = \Delta_B \bar{L}^3 k_F^{-2}(-\sin \bar{L} + \sinh \bar{L})$$

$$\Delta_B = \frac{1}{1 - \cos \bar{L} \cosh \bar{L}}$$

$$\bar{L} = k_F L$$

### 3.1.2 Transfer Matrix

The left and right sides of the Euler-Bernoulli beam can be correlated with a transfer matrix. In Fig. 3.1, we can write the different force and displacement vectors for the left-hand side and right-hand side. The force vectors for the left-hand side $f_L$ and the right-hand side $f_R$ are:

$$\begin{aligned} f_L &= \left\{ \begin{array}{cc} Q(0) & M(0) \end{array} \right\}^T \\ f_R &= \left\{ \begin{array}{cc} Q(L) & M(L) \end{array} \right\}^T \end{aligned} \quad (3.11)$$

The displacement vectors for the left-hand side $d_L$ and the right-hand side $d_R$ are expressed as:

$$\begin{aligned} d_L &= \left\{ \begin{array}{cc} W(0) & \theta(0) \end{array} \right\}^T \\ d_R &= \left\{ \begin{array}{cc} W(L) & \theta(L) \end{array} \right\}^T \end{aligned} \quad (3.12)$$

The dynamic stiffness matrix is expressed in the following format as shown by Kumar *et al.* (2023)

$$\begin{bmatrix} s_{LL} & s_{LR} \\ s_{RL} & s_{RR} \end{bmatrix} \begin{Bmatrix} d_L \\ d_R \end{Bmatrix} = \begin{Bmatrix} f_L \\ f_R \end{Bmatrix} \quad (3.13)$$

The force-displacement connection can be converted into an output-input relationship

as follows:

$$\begin{Bmatrix} d_R \\ f_R \end{Bmatrix} = T(\omega) \begin{Bmatrix} d_L \\ f_L \end{Bmatrix} \qquad (3.14)$$

where $T(\omega)$ is the transfer matrix. It can be represented as:

$$T(\omega) = \begin{bmatrix} -s_{LR}^{-1} s_{LL} & -s_{LR}^{-1} \\ s_{RL} - s_{RR} s_{LR}^{-1} s_{LL} & -s_{RR} s_{LR}^{-1} \end{bmatrix} \qquad (3.15)$$

In the case of a periodic metamaterial beam, the transfer matrix is computed for the single-unit cell (which is the building block for periodic structure) of the periodic beam. The dispersion relationship can then be calculated by combining the transfer matrix with Bloch-Floquet's theorem. This has been clearly explained in Chapter 4.

# CHAPTER 4

# COUPLED ACOUSTIC BLACK HOLE & LOCAL RESONATOR: ANALYTICAL MODEL

## 4.1 INTRODUCTION

Periodic structures like metamaterials can be employed to vibration attention due to their novelty in specific frequency filtering. This property has been investigated by Mead (1973); Orris and Petyt (1974); Mace (1984). Brillouin (1953) has observed that periodic structures, called phononic metamaterial, show propagation and attenuation bands at specific frequency regions. A novel sonic bandgap material owing to local resonances has been demonstrated experimentally and theoretically by Liu *et al.* (2000). Attenuation bands are generated in these metamaterials due to phenomena such as Bragg-type scattering (BS), Local Resonance (LR), or their coupling in general, as discussed by Chang *et al.* (2018). Consequently, researchers have shown considerable interest in lightweight periodic structures for mechanical and aerospace applications. Wave propagation in beam structures with periodic ABH has been investigated by an energy-based method by Tang and Cheng (2017*a*). Embedding multiple ABH in a structure has been shown by Gao *et al.* (2022); Deng *et al.* (2019) to attenuate the mid-high frequency vibrations. Similarly, the periodic local resonators are found to attenuate the vibration at low frequencies.

Many researchers have shown interest in the coupled effects of metamaterials to widen the attenuation characteristics. Hu *et al.* (2017) have studied coupled local resonators to achieve broadband vibration attenuation in the acoustic metamaterial. Dwivedi *et al.* (2022) have studied the coupled negative mass and negative stiffness resonators and observed the merging effect in bandgaps. Mondal *et al.* (2023) have studied the coupled local resonators for flexural and torsional vibration attenuation simultaneously.

In this part, coupled ABHLR metamaterial is studied for low to high-frequency vibration attenuation. An exact closed-form solution is derived using the second-order power-law profile based on Euler-Bernoulli's beam theory. From this, the dynamic stiffness matrix, integrating local resonance effects, is derived using the spectral element method, as discussed by Lee (2009). The assembling techniques of the spectral element matrix are the same as the traditional finite element method. Dispersion diagrams are generated by combining Bloch's theorem with the ABHLR transfer matrix. Finally, the frequency response function is analysed from the stiffness matrix of the finite ABHLR beam, and the results are validated with the dispersion analysis of the unit cell. The frequency response is also validated using COMSOL Multiphysics (2018) finite element simulations.

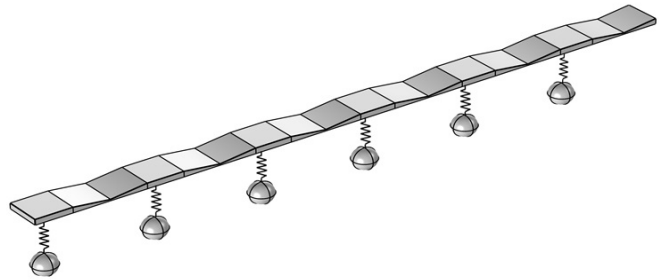

Figure 4.1: COMSOL finite ABHLR beam diagram with six unit cells.

## 4.2 DYNAMIC STIFFNESS METHOD FOR ABHLR CELL

Fig. 4.1 represents the finite coupled ABHLR beam with six unit cells modelled by COMSOL Multiphysics. Fig. 4.2 shows the representative diagram of coupled ABHLR unit cell with proper dimensions. This coupled unit cell is divided into four sub-units.

1. Uniform beam with LR sub-unit (length $l_1$): connected with nodes 1 and 2.

2. Uniform beam sub-unit (length $l_2$): connected with nodes 2 and 3.

3. ABH: Downward wedge sub-unit (length $l_3$): connected with nodes 3 and 4.

4. ABH: Upward wedge sub-unit (length $l_4$): connected with nodes 4 and 5.

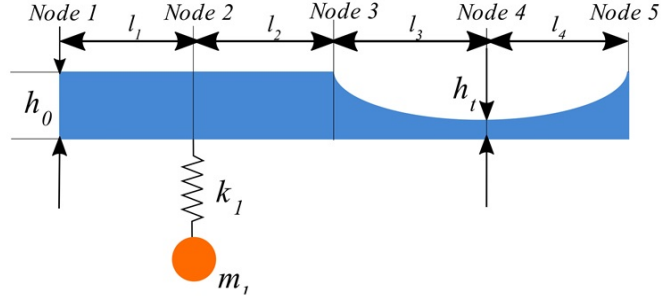

Figure 4.2: Representative diagram of coupled ABHLR unit cell. The figure shows four sub-unit cells with proper dimensions.

The exact closed-form solutions are derived for each sub-unit based on Euler-Bernoulli's beam theory. From these, dynamic stiffness (DS) (Leung (2012)) matrices are developed using the spectral element method for each sub-unit. Finally, the dynamic stiffness matrix of coupled ABHLR unit is obtained by combining the DS matrices of each sub-unit.

### 4.2.1 ABH: Downward wedge sub-unit

The governing equation of Euler- Bernoulli's beam for a particular frequency $\omega$, with $\omega \in \{\omega_j\}, j = 0, 1, ...n$ can be written as

$$\frac{d^2}{dx^2}\left(EI(x)\frac{d^2W}{dx^2}\right) - \rho A(x)\omega^2 W = 0 \qquad (4.1)$$

where material parameters $\rho$, and $E$ denote the density and Young's modulus of the beam, respectively. $W(x)$ represents the amplitude of deflection at a specific frequency. The second moment of inertia of any section at distance $x$, with $x \in (0, l)$ of each sub-unit, is represented as $I(x) = \frac{bh(x)^3}{12}$, where $l$ is the length of respective sub-unit and width $b$ is fixed throughout the unit cell. The geometric parameters, $h(x)$ and $A(x)$, are the thickness and cross-sectional area corresponding to *x* distance of each sub-unit, respectively.

For the ABH section, the thickness varies according to the power law profile along the length of the sub-unit. To propose a simplified formulation, the parabolic power-law profile is considered throughout this study. The dynamic stiffness for the higher-order profile can be formulated in a similar manner. For the downward wedge section, the thickness varies along the x-direction in a parabolic fashion as

$$h(x) = h_0 \left(1 - \frac{x}{x_0}\right)^2 \tag{4.2}$$

where $h_0$ is the thickness of the beam at the junction of uniform and downward wedge sections at $x = 0$. $x_0$ is the length, where, ideally, taper thickness becomes zero. The length of this portion $l_3 = (1 - \sqrt{r_0})x_0$, where $r_0 = \frac{h_t}{h_0}$. $h_t$ is the thickness at the junction between the downward and upward wedge sections at $x = l_3$. The exact solution of the downward wedge section has been derived in Appendix A.

The exact solution of the downward wedge section corresponding to the above parabolic thickness can be written as discussed by Lee and Jeon (2019)

$$\begin{aligned} W(x) &= a_1 \left(1 - \frac{x}{x_0}\right)^{c_1} + a_2 \left(1 - \frac{x}{x_0}\right)^{c_2} + a_3 \left(1 - \frac{x}{x_0}\right)^{c_3} + a_4 \left(1 - \frac{x}{x_0}\right)^{c_4} \\ &= \sum_{i=1}^{4} a_i \left(1 - \frac{x}{x_0}\right)^{c_i} \end{aligned} \tag{4.3}$$

where $c_1, c_2, c_3$ and $c_4$ are the function of the particular frequency $\omega$. These can be represented by

$$\begin{aligned} c_1 &= -\frac{3}{2} - \frac{1}{2}\sqrt{17 - 4\sqrt{4 + \lambda}} \\ c_2 &= -\frac{3}{2} - \frac{1}{2}\sqrt{17 + 4\sqrt{4 + \lambda}} \\ c_3 &= -\frac{3}{2} + \frac{1}{2}\sqrt{17 - 4\sqrt{4 + \lambda}} \\ c_4 &= -\frac{3}{2} + \frac{1}{2}\sqrt{17 + 4\sqrt{4 + \lambda}} \end{aligned} \tag{4.4}$$

In the Eq. 4.4, $\lambda = 12x_0^4 \rho \omega^2 / \left(E h_0^2\right)$ and $a_1, a_2, a_3, a_4$ are the coefficients which can be written as $A = [a_1 \ a_2 \ a_3 \ a_4]^T$. The DS matrix is obtained by formulating the spectral element matrix of this sub-unit cell. The nodal displacement field can be defined as $U_3 = H^d A$, and the nodal force can be constructed as $F_3 = R^d A$ for the downward-wedge section. The nodal displacement and force can be represented as

$$U_3 = [W(0) \ W'(0) \ W(l_3) \ W'(l_3)]^T = \sum_{i=1}^{4} a_i H^d_i$$

$$F_3 = [-Q(0) \ -M(0) \ Q(l_3) \ M(l_3)]^T = \sum_{i=1}^{4} a_i R^d_i$$

(4.5)

where $H^d_i$ and $R^d_i$ are $4 \times 1$ vectors. In the Eq. 4.5, $W$, $W'$, $M$ and $Q$ represent transverse displacement, degree of rotation, bending moment and shear force, respectively. The bending moment and shear force can be written as

$$Q = -\frac{d}{dx}\left(EI(x)\frac{d^2W}{dx^2}\right)$$

$$M = EI(x)\frac{d^2W}{dx^2}$$

(4.6)

From Eq. 4.5 and Eq. 4.6, $H^d_i$ and $R^d_i$ vectors with respect to the downward wedge section can be obtained as

$$H^d_i = \begin{bmatrix} 1 \\ -\frac{c_i}{x_0} \\ \left(1 - \frac{l_3}{x_0}\right)^{c_i} \\ -\frac{c_i}{x_0}\left(1 - \frac{l_3}{x_0}\right)^{c_i-1} \end{bmatrix}$$

(4.7)

$$R^d_i = \frac{Ebh_0^3}{12x_0^3} \begin{bmatrix} -\left(c_i^3 + 3c_i^2 - 4c_i\right) \\ -c_i(c_i - 1)x_0 \\ \left(c_i^3 + 3c_i^2 - 4c_i\right)\left(1 - \frac{l_3}{x_0}\right)^{c_i+3} \\ x_0 c_i (c_i - 1)\left(1 - \frac{l_3}{x_0}\right)^{c_i+4} \end{bmatrix}$$

(4.8)

From the Eq. 4.7 and Eq. 4.8, the $H^d$ matrix and $R^d$ matrix are generally getting in the following way, as given below

$$H^d = \begin{bmatrix} H^d_1 & H^d_2 & H^d_3 & H^d_4 \end{bmatrix}$$
$$R^d = \begin{bmatrix} R^d_1 & R^d_2 & R^d_3 & R^d_4 \end{bmatrix} \quad (4.9)$$

The DS matrix $K_{ABHd}$ are represented by the force-displacement relationship $F_3 = K_{ABHd}U_3$. After eliminating the constant $A$ from Eq. 4.5, we can get

$$K_{ABHd} = R^d {H^d}^{-1} = \begin{bmatrix} K_{31} & K_{32} \\ K_{33} & K_{34} \end{bmatrix} \quad (4.10)$$

The $4 \times 4$ matrix $K_{ABHd}$ represents the DS matrix of the downward wedge ABH sub-unit. $K_{31}, K_{32}, K_{33}, K_{34}$ are all $2 \times 2$ matrices, defined to represent four quadrants of the dynamic stiffness matrix.

### 4.2.2 ABH: Upward wedge sub-unit

For the upward wedge-ABH sub-unit, the thickness increases along the x-direction in a parabolic manner, as given below

$$h(x) = h_t \left(1 + \frac{x}{x_t}\right)^2 \quad (4.11)$$

where $h_t$ is the junction of the downward and upward wedge-ABH sections at $x = 0$. The length of this sub-unit can be expressed by $l_4 = (\sqrt{r_t} - 1)x_t$, where $r_t$ denotes $r_t = \frac{h_0}{h_t}$. $h_0$ is the thickness at the end of the upward wedge section at $x = l_4$. The exact solution of the upward wedge section corresponding to the above parabolic equation can be written as discussed by Lee and Jeon (2019).

$$W(x) = b_1\left(1 + \frac{x}{x_t}\right)^{c_1} + b_2\left(1 + \frac{x}{x_t}\right)^{c_2} + b_3\left(1 + \frac{x}{x_t}\right)^{c_3} + b_4\left(1 + \frac{x}{x_t}\right)^{c_4} \quad (4.12)$$

In the upward wedge section, the constants $c_1, c_2, c_3, c_4$ have the same relations as Eq. 4.4. But the value of $\lambda$ is different. It can be expressed as $\lambda = 12x_t^4 \rho \omega^2/(E h_t^2)$. The parameters, $b_1, b_2, b_3, b_4$ are the coefficients which can be written as $B = [b_1\ b_2\ b_3\ b_4]^T$. Similarly, the nodal displacement field can be defined as $U_4 = H^u B$, and the nodal force can be constructed as $F_4 = R^u B$ for the upward wedge section. The vectors $H^u{}_i$ and $R^u{}_i$ can be solved in a similar way like Eq. 4.7 and Eq. 4.8. Finally, the $H^u$ and $R^u$ matrices can be derived, as given below

$$H^u{}_i = \begin{bmatrix} 1 \\ \frac{c_i}{x_t} \\ (1 + \frac{l_4}{x_t})^{c_i} \\ \frac{c_i}{x_t}(1 + \frac{l_4}{x_t})^{c_i-1} \end{bmatrix} \quad (4.13)$$

$$R^u{}_i = \frac{Ebh_t^3}{12x_t^3} \begin{bmatrix} (c_i^3 + 3c_i^2 - 4c_i) \\ -c_i(c_i-1)x_t \\ -(c_i^3 + 3c_i^2 - 4c_i)(1 + \frac{l_4}{x_t})^{c_i+3} \\ x_t c_i (c_i - 1)(1 + \frac{l_4}{x_t})^{c_i+4} \end{bmatrix} \quad (4.14)$$

The DS matrix $K_{ABHu}$ are represented by the force-displacement relationship $F_4 = K_{ABHu} U_4$. After eliminating the constant $B$, we can get

$$K_{ABHu} = R^u H^{u-1} = \begin{bmatrix} K_{41} & K_{42} \\ K_{43} & K_{44} \end{bmatrix}. \quad (4.15)$$

The $4 \times 4$ matrix $K_{ABHu}$ represents the DS matrix of the upward wedge ABH sub-unit. ,$K_{41}, K_{42}, K_{43}, K_{44}$ are all $2 \times 2$ matrices, defined to represent four quadrants of the dynamic stiffness matrix.

### 4.2.3 Uniform beam sub-unit

The general equation of Euler- Bernoulli's beam of the uniform cross-section for a particular wave number $k$, with $k \in \{k_j\}$, $j = 0, 1, ...n$ can be written as

$$W(x) = d_1 e^{-ikx} + d_2 e^{-kx} + d_3 e^{+ikx} + d_4 e^{+kx} \qquad (4.16)$$

where the wave number for flexural wave mode can be defined by $k = \sqrt{\omega} \left(\frac{\rho A}{EI}\right)^{1/4}$. $A$ represents the constant area of the uniform beam section. $D = [d_1 \; d_2 \; d_3 \; d_4]^T$ is the coefficient vector.

The DS matrix of the uniform beam section $K_u$ can be obtained by applying the spectral element method using the force-displacement relationship $F_2 = K_u U_2$. The DS matrix $K_u$ can be represented as

$$K_u = \frac{EI}{l_2^3} \begin{bmatrix} K_{B11} & K_{B12} & K_{B13} & K_{B14} \\ K_{B12} & K_{B22} & K_{B23} & K_{B24} \\ K_{B13} & K_{B23} & K_{B33} & K_{B34} \\ K_{B14} & K_{B24} & K_{B34} & K_{B44} \end{bmatrix} = \begin{bmatrix} K_{21} & K_{22} \\ K_{23} & K_{24} \end{bmatrix} \qquad (4.17)$$

where $K_{21}, K_{22}, K_{23}, K_{24}$ are all $2 \times 2$ matrices, defined to represent four quadrants of the dynamic stiffness matrix. The elements of Eq. 4.17 matrix are discussed by Lee (2009). The elements of uniform beam stiffness $K_u$ are shown below

$$L = kl_2, \; c = \cos L, \; ch = \cosh L, \; s = \sin L, \; sh = \sinh L$$

$$m = \frac{1}{1 - cch}, \; K_{B11} = K_{B33} = mL^3(csh + sch)$$

$$K_{B22} = K_{B44} = mL^3 k^{-2}(-csh + sch)$$

$$K_{B12} = -K_{B34} = mL^3 k^{-1} ssh \qquad (4.18)$$

$$K_{B13} = -mL^3(s + sh)$$

$$K_{B14} = -K_{B23} = mL^3 k^{-1}(-c + ch)$$

$$K_{B24} = mL^3 k^{-2}(-s + sh)$$

### 4.2.4 Uniform beam with LR sub-unit

The local resonator attached to the beam can be considered a discrete spring-mass system. The general equation of discrete spring-mass for a particular frequency $\omega$, with $\omega \in \{\omega_j\}, j = 0, 1, ...n$ can be defined as shown by Hao *et al.* (2019)

$$\left(K - \omega^2 M\right) U_1 = F \tag{4.19}$$

where K, M, U, F are the stiffness matrix, mass matrix, displacement vector and force vector, respectively. These can be represented as

$$K = \begin{bmatrix} k_1 & -k_1 \\ -k_1 & k_1 \end{bmatrix} \quad M = \begin{bmatrix} 0 & 0 \\ 0 & m_1 \end{bmatrix}$$
$$U = \begin{bmatrix} v_b \\ v_s \end{bmatrix} \quad F = \begin{bmatrix} F_1 \\ 0 \end{bmatrix} \tag{4.20}$$

where $v_b$ is the transverse displacement of the beam at the resonator connecting point. $v_s$ is the displacement of the resonator mass $m_1$. The stiffness of the spring connected with the beam is $k_1$. $F_1$ is the force acting on the resonator connecting point of the uniform beam. It can be defined as $F_1 = k_{lr} v_b$. $k_{lr}$ represents the additional dynamic stiffness to the beam structure due to the local resonator. It has the following form

$$k_{lr} = k_1 - \frac{k_1^2}{k_1 - \omega^2 m_1} \tag{4.21}$$

The DS matrix for the local resonator can be written as

$$\boldsymbol{D} = \begin{bmatrix} k_{lr} & 0 \\ 0 & 0 \end{bmatrix} \tag{4.22}$$

The DS matrix for the uniform beam part is shown in the previous section. The length of the beam is $l_1$. Four quadrants of the stiffness matrix (uniform beam section) are denoted by $K_{11}, K_{12}, K_{13}, K_{14}$. The dynamic stiffness matrix of the uniform beam with a local resonator can be represented as

$$K_{ul} = \begin{bmatrix} K_{11} & K_{12} \\ K_{13} & K_{14} + D \end{bmatrix} \tag{4.23}$$

## 4.3 ABHLR TRANSFER MATRIX AND DISPERSION RELATIONSHIP

Dispersion relations are generated to get wave propagation characteristics for infinite periodic structures. These relations provide the real and imaginary values of wave numbers with their frequency. The zero values of imaginary wave numbers signify the wave propagation band, whereas non-zero values indicate the presence of an attenuation band. The transfer matrix of the unit cell is combined with Bloch's theorem to generate a dispersion or band diagram. The transfer matrix relates the left and right side nodes of the unit cell.

The transfer matrix of the unit cell can be derived from the spectral element or dynamic stiffness matrix. It can be written as

$$T = \begin{bmatrix} -K_{LR}^{-1}K_{LL} & -K_{LR}^{-1} \\ K_{RL} - K_{RR}K_{LR}^{-1}K_{LL} & -K_{RR}K_{LR^{-1}} \end{bmatrix} \tag{4.24}$$

where $K$ is the dynamic stiffness matrix of the unit cell. It can be represented by

$$K = \begin{bmatrix} K_{LL} & K_{LR} \\ K_{RL} & K_{RR} \end{bmatrix} \tag{4.25}$$

where $K_{LL}, K_{LR}, K_{RL}$ and $K_{RR}$ are four quadrants of the dynamic stiffness matrix $K$.

For the coupled ABHLR cell, the transfer matrix of each sub-unit cell is derived from its respective dynamic stiffness matrix. Finally, the transfer matrix of the coupled ABHLR unit cell can be computed as

$$T_{ABHLR} = T_4 T_3 T_2 T_1 \tag{4.26}$$

where $T_i (i = 1, 2, 3, 4)$ represents the transfer matrix of $i^{th}$ sub-unit of the coupled ABHLR cell.

According to Bloch-Floquet's theorem, the relationship between successive unit cells can be written as

$$\psi_n(1) = e^{-iZ} I \psi_{n-1}(1) \rightarrow \psi_n(1) = e^{-iZ} I \psi_n(5) \tag{4.27}$$

$\psi_n(i)$ denotes the state vector at $i^{th}$ node of $n^{th}$ unit cell and $Z$ denotes the wave-number of coupled ABHLR unit cell. The dispersion diagram can be computed by combining the transfer matrix with the state relation between $1^{st}$ and $5^{th}$ nodes for the coupled ABHLR cell. This relation can be written as

$$\begin{aligned} \psi_n(1) &= T_{ABHLR} \psi_n(5) \\ \left| T_{ABHLR} - e^{-iZ} \cdot I \right| \psi_n(5) &= 0 \\ \rightarrow e^{-iZ} &= \underbrace{\text{eig}(T_{ABHLR})}_{q} \rightarrow Z = -i \ln(q) \end{aligned} \tag{4.28}$$

The minimum non-zero value of the imaginary axis ($Im(Z)$) represents the attenuation rate of non-propagating waves as discussed by Wang and Wang (2013); Banerjee (2020).

## 4.4 FINITE ABHLR BEAM: FRF

The frequency response function (FRF) is a function used to quantify the response of a finite structure under an excitation force. In this study, the finite beam is considered with 30 ABH or ABHLR unit cells. It is maintained with free-free boundary conditions at both ends. The FRF studies are obtained at one end (output end) of the beam under an excitation force at the other end (input end). The frequency response studies are analysed to validate dispersion bandgaps for finite beams. The frequency response function generally generates a series of peaks in the propagation band and a drop in the attenuation band.

The dynamic stiffness $K_{ABHLR}$ of the coupled ABHLR cell can be derived by assembling the stiffness of each sub-unit. It can be written as

$$K_{ABHLR} = \begin{bmatrix} K_{11} & K_{12} & 0 & 0 & 0 \\ K_{13} & K_{14} + D + K_{21} & K_{22} & 0 & 0 \\ 0 & K_{23} & K_{24} + K_{31} & K_{32} & 0 \\ 0 & 0 & K_{33} & K_{34} + K_{41} & K_{42} \\ 0 & 0 & 0 & K_{43} & K_{44} \end{bmatrix} \quad (4.29)$$

The global stiffness matrices are developed by combining the dynamic stiffness of 30 ABH or ABHLR unit cells. The frequency response function can be obtained from the global stiffness matrix of the finite beam. The logarithmic output responses of a finite beam under unit harmonic input excitation are performed on a decibel scale, which is called the transmissibility of the beam.

This analytical model has been validated by COMSOL Multiphysics finite element simulations in Chapter 5. Numerical results are also discussed in that chapter.

# CHAPTER 5

# COUPLED ACOUSTIC BLACK HOLE & LOCAL RESONATOR: NUMERICAL RESULTS

In this Chapter, we have discussed the validation of the analytical model using COMSOL Multiphysics finite element simulation software. Later, numerical results have been discussed to find the insights due to the coupling effect. We have explained clearly the physics behind the merging effect of acoustic black holes and local resonators. Finally, conclusions are drawn for the coupled effect in metamaterial beam for flexural vibration attenuation.

## 5.1 FINITE ELEMENT VALIDATION OF DYNAMIC STIFFNESS METHOD

The dynamic stiffness method must be validated before the investigation of wave propagation characteristics. A finite element simulation framework is developed by commercial software COMSOL Multiphysics. FRF studies are performed for finite beams under unit harmonic perturbation. For this purpose, finite beams with 6 unit cells are considered to reduce the computational time of finite element simulations. The finite element mesh of finer size is used to converge the results. Finally, the transmissibility responses are observed by using the following formula

$$T = 20 \log \frac{d_{out}}{d_{in}} \tag{5.1}$$

where $d_{out}$, $d_{in}$, and $T$ denote output displacement amplitudes, input displacement amplitudes and transmissibility response, respectively. Fig. 5.1 represents the transmissibility response of the finite ABH beam, and Fig. 5.2 shows the transmissibility response of the finite ABHLR beam (with spring stiffness $k_1=10^5$ N/m and resonator mass $m_1=0.032$ kg). These results provide a satisfactory match between the dynamic stiffness method and COMSOL simulation at the frequency range of interest.

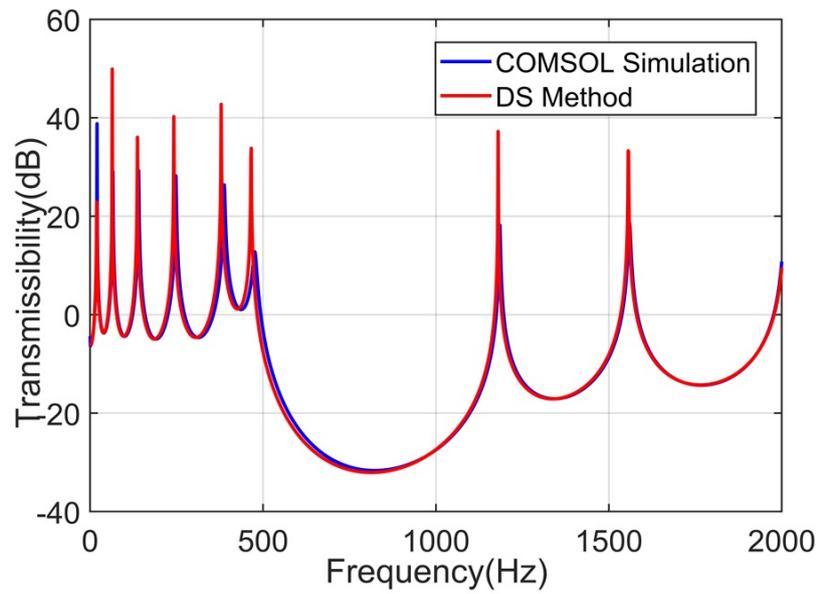

Figure 5.1: Validation of dynamic stiffness method.Transmissibility plot of the finite beam with 6 ABH units

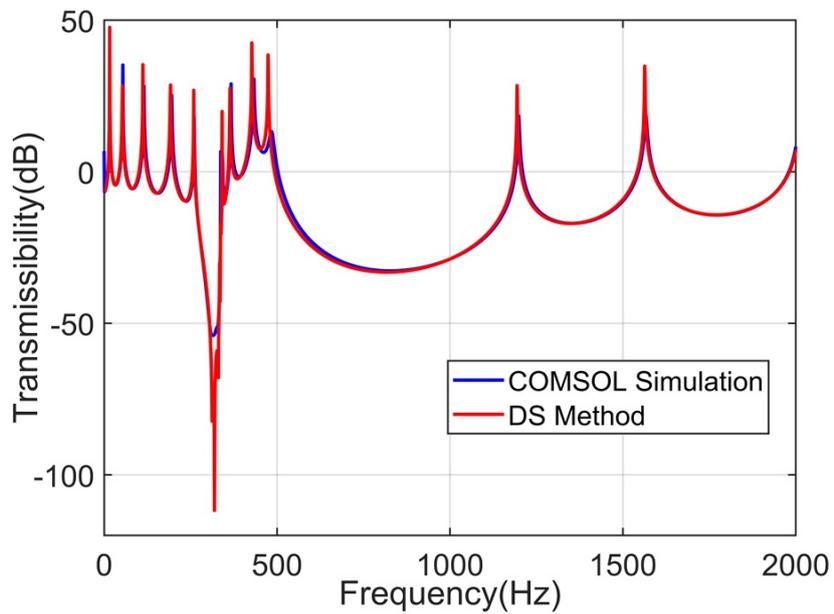

Figure 5.2: Validation of dynamic stiffness method. Transmissibility plot of the finite beam with 6 ABHLR units.

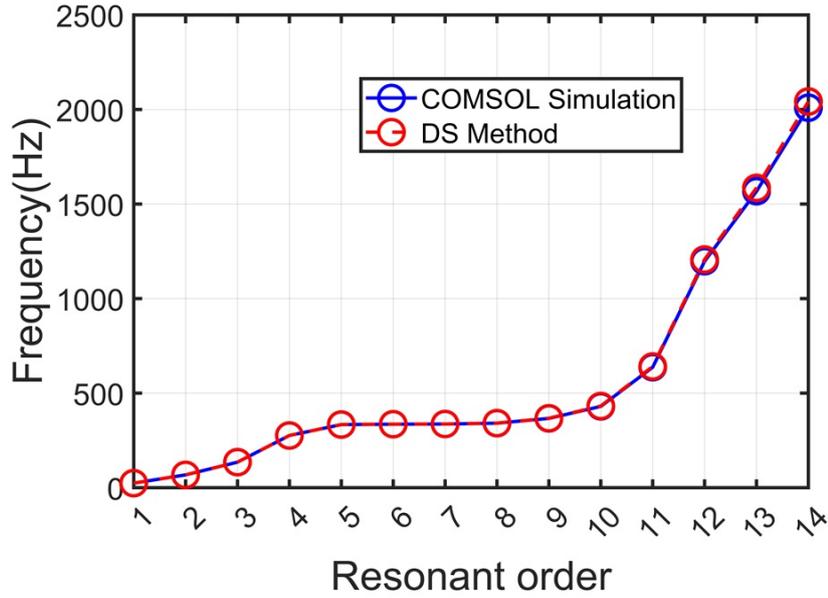

Figure 5.3: Eigenfrequencies of finite ABHLR beam with the first 14 modes are shown. The red and blue lines represent the eigenfrequencies obtained from the DS method and COMSOL simulations, respectively.

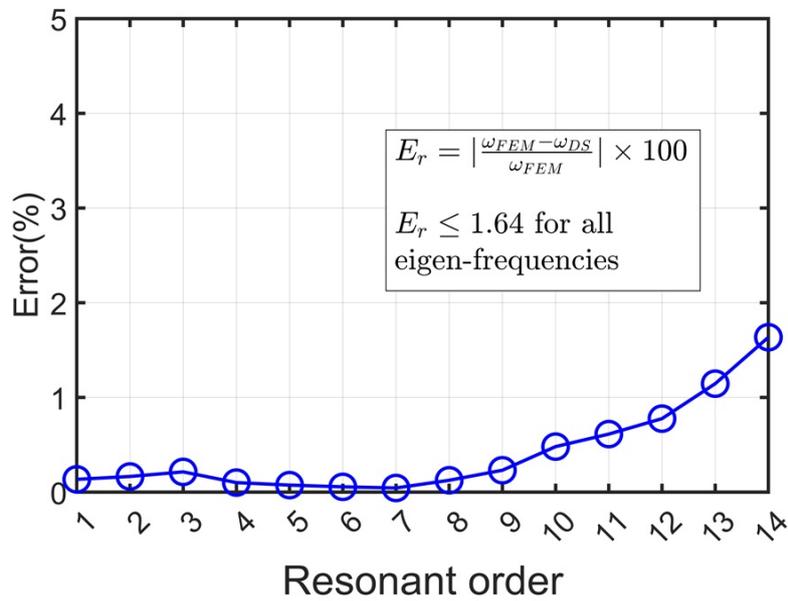

Figure 5.4: This figure represents the errors in calculating each eigenfrequency obtained from the DS method.

Further, the natural frequencies of the finite ABHLR beam are obtained by equating the determinant of the global stiffness matrix to zero ($|K| = 0$). In Fig. 5.3, the natural frequencies of finite ABHLR beam are compared with the COMSOL simulation results, and the errors in obtaining natural frequencies from the DS method are shown in Fig. 5.4. The formula used to calculate the error is

$$E_r = |\frac{\omega_{FEM} - \omega_{DS}}{\omega_{FEM}}| \times 100\% \quad (5.2)$$

where $\omega_{FEM}$ and $\omega_{DS}$ denote the natural frequency obtained from COMSOL simulations and DS method, respectively. The percentage of error value is represented by $E_r$, which is limited to [0.04,1.64] %. As maximum value is so small that two curves in Fig. 5.2 can be considered as coincided. These results finally indicate the reliability of the dynamic stiffness method.

Table 5.1: Geometrical and material Parameters of the ABH and coupled ABHLR unit cell

| Geometry | Material |
|---|---|
| $h_0 = 5mm$ | $\rho = 7850 kg/m^3$ |
| $h_t = 1mm$ | $E = 200$ GPa |
| $b = 30mm$ | |
| $l_1 = l_2 = 15mm$ | |
| $l_3 = l_4 = 30mm$ | |

## 5.2 DISPERSION AND FRF RESULTS

Dispersion characteristic depends on the geometrical and material properties of the unit cell. It reflects wave propagation characteristics of the infinite periodic unit cell. The concise dispersion diagrams (real and imaginary axis) are shown for the ABH and coupled ABHLR unit cell. The minimum non-zero value of the imaginary axis (*Im(Z)*) represents the attenuation rate of non-propagating waves as discussed by Wang and Wang (2013); Banerjee (2020).

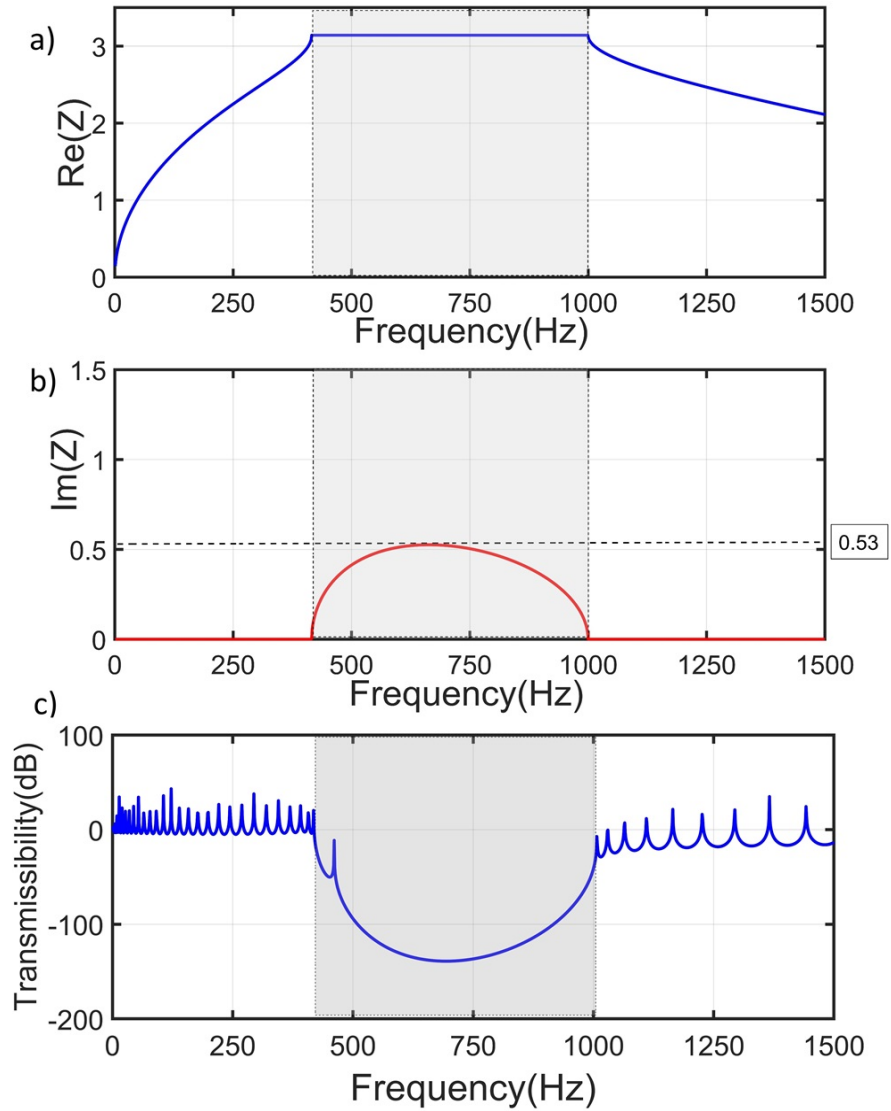

Figure 5.5: Concise dispersion diagram of ABH unit cell (Without local resonator): a) Real axis, b) Imaginary axis (Attenuation band = 416 to 1000 Hz). Fig. c is the Transmissibility plot of the finite beam with 30 ABH units.

The frequency response studies represent the applicability of the infinite unit structures in the finite range. The attenuation level of the finite structure increases with the addition of the unit cell. It is noteworthy that dispersion analysis of a unit cell as an alternative to FRF study for the entire beam can significantly reduce computational complexity. However, it is essential to determine the minimum number of unit cells within the finite beam required for achieving convergence between the dispersion and FRF results. In this study, finite beams with 30 unit cells have been employed to ensure convergence with the results obtained through dispersion analysis.

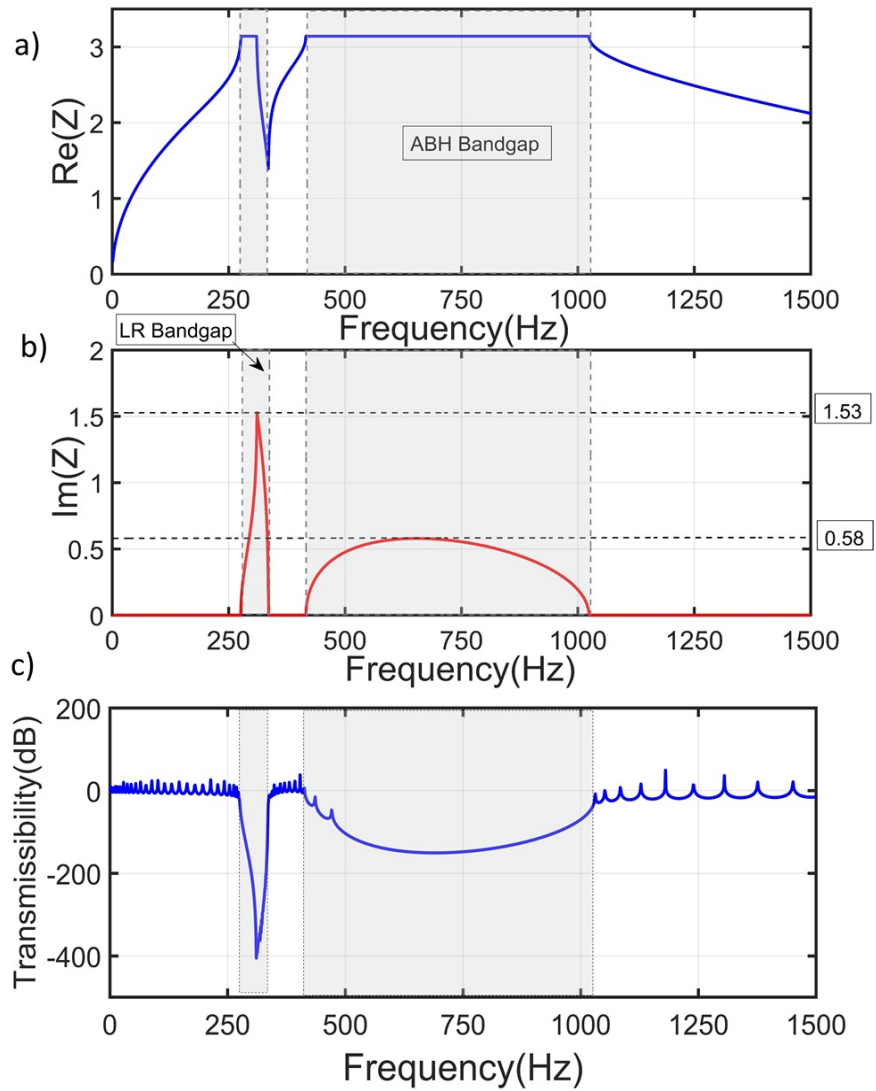

Figure 5.6: Concise dispersion diagram of coupled ABHLR unit cell (With local resonator): a) Real axis, b) Imaginary axis, (LR attenuation band = 276 to 336 Hz, ABH attenuation band = 404 to 1030 Hz). Fig. c is the Transmissibility plot of the finite beam with 30 ABHLR units.

Fig. 5.5 represents the dispersion diagram of the ABH unit cell and the transmissibility diagram of the finite ABH beam. The dispersion diagram shows an attenuation band between 416 − 1000 Hz with a peak attenuation rate $Im(Z)$ = 0.53, whereas the transmissibility diagram represents the same bandgap with a maximum transmissibility of $-140$ dB. The concise dispersion and transmissibility diagrams of the coupled ABHLR unit cell are shown in Fig. 5.6. In addition to the above parameters, a mass-spring system with spring stiffness $K_1 = 10^5$ N/m and mass $m_1 = 0.032$ kg is attached. For the coupled ABHLR cell, the ABH attenuation band between $404 - 1030$ Hz with a high attenuation rate compared to the ABH cell is found in both dispersion and transmissibility diagrams. The dispersion diagram represents the maximum attenuation rate $Im(Z) = 0.58$, corresponding to maximum transmissibility $-150$ dB at the transmissibility diagram for ABH bandgap. Along with the ABH bandgap, coupled ABHLR presents an LR attenuation band between $276 - 336$ Hz with the maximum attenuation rate $Im(Z) = 1.53$. It shows the same LR attenuation band with maximum transmissibility of $-410$ dB at the transmissibility diagram. The LR attenuation band exhibits almost 170 % higher attenuation rate than the ABH attenuation band.

## 5.3 EFFECT OF COUPLING ABH AND LR

This section presents the coupled ABHLR effect due to the overlapping of acoustic black hole and local resonance phenomena. Within the entire overlapping zone, both constructive and destructive interference emerge simultaneously owing to the superposition of local resonance and acoustic black hole effect. The wave propagation band is developed due to constructive interference, while the attenuation band results from destructive interference. The local resonance effect interferes with the wave-trapping property of the acoustic black hole, yielding specific wave propagation characteristics within this zone.

To understand the new pass-band formation mechanism within two merged LR attenuation bands, a linear momentum analysis of the unit cell has been conducted by Zhu *et al.* (2014). It is known that when the total linear momentum of the entire system (host beam + resonator) is out of phase with the host beam, vibration cannot be transmitted through the beam, and a bandgap is generated. Later, Xiao *et al.* (2013); Sharma and Sun (2016); Cenedese *et al.* (2021) investigated the interaction between LR and Bragg bandgaps in periodic unit cells. In the merging region, the scattered waves generated by Bragg's scattering effect interact with the vibration of local resonance. If the total linear momentum of the system is in phase with the host beam, a passband is generated. In other words, two attenuation bands are separated by a pass-band due to the merging effect, as discussed by Sharma and Sun (2016).

For the ABH unit cell (without LR), if the characteristic length of the periodic ABH unit coincides with the wavelength, then the scattered waves will be out-of-phase with propagating waves and result in destructive interference. Therefore, an attenuation band is generated, which is the same as Bragg's scattering effect. For the LR unit cell (without ABH), when the frequency of propagating waves is near the resonance frequency of the local resonator, then LR oscillates out of phase with the host beam. This will result in an attenuation band, as discussed by Deymier (2013).

For the coupled ABHLR unit cell, within the ABH bandgap region, the linear momentum of the coupled cell stays out-of-phase with the host beam. However, when the LR bandgap overlaps with the ABH bandgap, the LR effect compels the linear momentum of the coupled cell to be in phase with the host beam in the overlapped region. This effect results in constructive interference, and a pass-band is generated. However, the remaining areas of the ABH bandgap region show destructive interference. The following subsection discusses three distinct characteristics of the coupled ABHLR cell.

### 5.3.1 Merged band: High attenuation rate

Fig. 5.7 illustrates a comparison of dispersion diagrams among the coupled ABHLR, ABH and LR unit cells. To facilitate a more effective comparison, the reduced mass of the engraved ABH is equated to the resonator mass in the coupled ABHLR cell. As a result, the existing mass of the ABHLR cell remains consistent with that of the original beam, without the inclusion of LR and ABH. Furthermore, the stiffness of the spring has been increased to $5 \times 10^5$ N/m to analyse the characteristics within the overlapping zone.

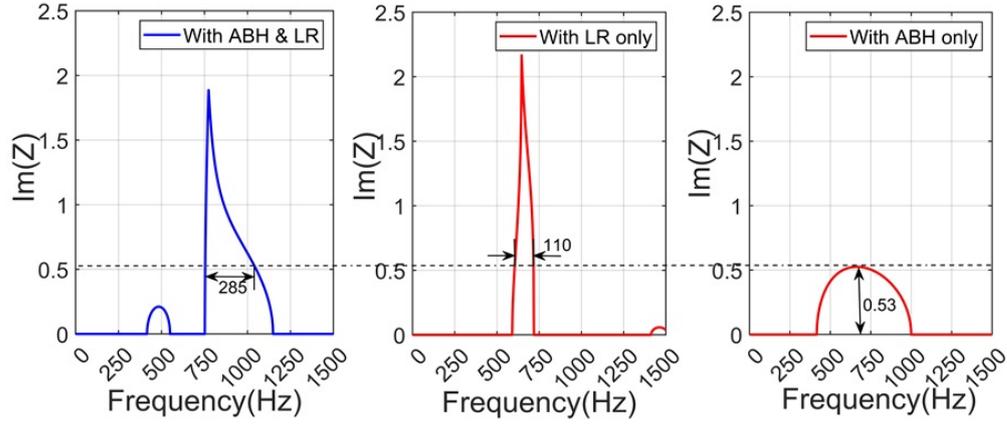

Figure 5.7: Comparison of attenuation band of the LR, ABH and coupled ABHLR unit cell. The attenuation bandwidths with higher attenuation, compared to the peak attenuation rate of the ABH cell are shown for the LR and coupled ABHLR unit cells.

In the coupled ABHLR cell, two distinct attenuation bands are observed due to destructive interference. An attenuation bandwidth of 285 Hz with higher attenuation, compared to the peak attenuation of the ABH cell ($Im(Z) = 0.53$) is found for the coupled cell. While the LR cell exhibits a 110 Hz higher bandwidth in comparison to the peak attenuation of the ABH cell. The coupled ABHLR cell establishes a 160% broader bandwidth with higher attenuation than the LR cell. Moreover, the simultaneous interaction of ABH and LR effects within the overlapping zone gives rise to a substantial outcome. This coupled effect can enhance the attenuation capabilities of the uncoupled local resonator or acoustic black hole configurations.

### 5.3.2 Merged band: Double-peak phenomenon

Fig. 5.8 represents a comparison of attenuation characteristics between the ABH and coupled ABHLR cells. The dispersion characteristic along the imaginary axis is plotted against a spring stiffness of $1.75 \times 10^5$ N/m. A double peak at a single attenuation band is observed due to the destructive interference. Bhatt and Banerjee (2022) explored the double-peak phenomena arising from simultaneous negative mass and stiffness effects. In this study, the first peak emerges from the local resonator effect, while the second peak results from the acoustic black hole effect. An attenuation bandwidth of 236 Hz, featuring a minimum attenuation rate of 0.63 is found as a result of the double-peak phenomenon.

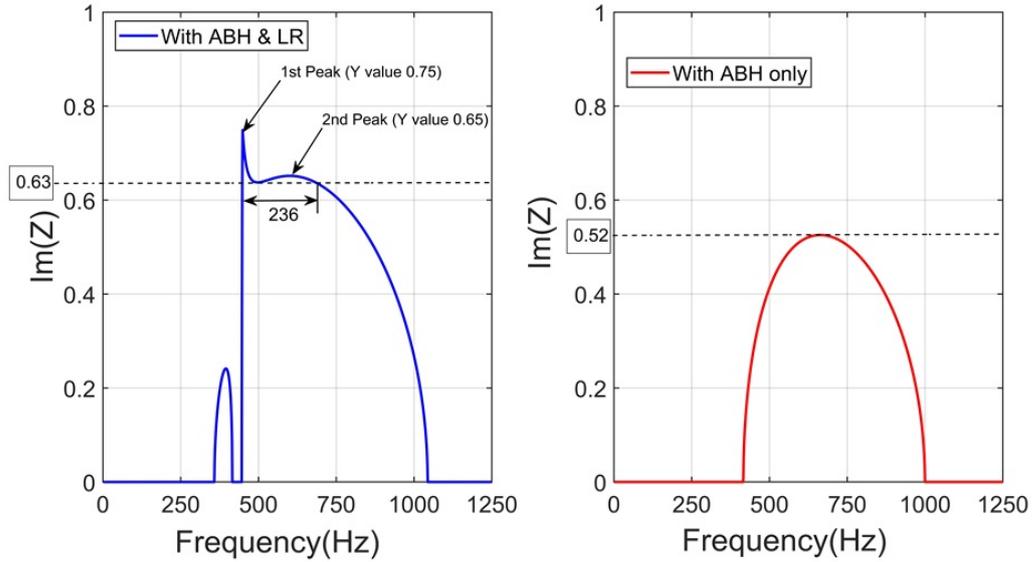

Figure 5.8: Comparison of attenuation characteristics ABH and ABHLR unit cell. A double peak at a single attenuation band is shown for the coupled ABHLR cell. Two peaks are observed due to the simultaneous interaction of LR and ABH effect.

This phenomenon becomes evident through the alteration of the LR band within the ABH bandgap or overlapping zone. The emergence of the double peak becomes visible when the cut-off frequency of the LR band aligns with the cut-on frequency of the ABH band. Beyond the cut-on frequency of ABH, the dominance of local resonance over the ABH peak becomes visible. Following a specific shift of the LR band, there is a transition where the double-peak phenomenon dissipates, giving rise to a single peak.

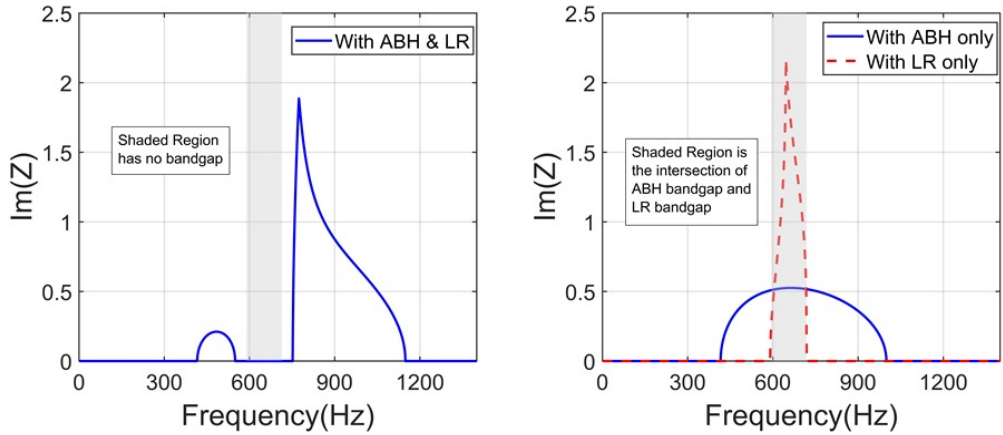

Figure 5.9: ABH-LR constructive interference phenomenon due to overlapping of the LR and ABH effect. The shaded region shows the attenuation band for individual ABH and LR effects but the pass-band for coupled ABHLR effects.

### 5.3.3 Merged band: Pass-band

Fig. 5.9 shows the dispersion diagrams resulting from the influences of the ABH, LR, and coupled ABHLR effects. This study presents the occurrence of pass-band due to constructive interference on the overlapping zone.

It is observed that the ABH effect shows an attenuation band between 416 to 1000 Hz, and the LR effect results in a bandgap from 590 to 720 Hz individually. The overlapping zone of these two bandgaps is shown with a grey-shaded region. A propagation or pass-band ranging from 590 to 720 Hz is found for coupled ABHLR cells, which is actually the attenuation band region for uncoupled ABH or LR effect. This phenomenon is consistently observed within the overlapping zone of ABH and LR bandgaps. It can be visualised comprehensively from the different parametric studies presented in Fig. 5.10. The coupling of ABH and LR bandgap merges but also produces the pass-band on the overlapping region due to constructive interference with each other. This phenomenon can be considered as a negative effect on bandgap formations. Therefore, the coupling has to be designed optimally to avoid constructive interference phenomenon.

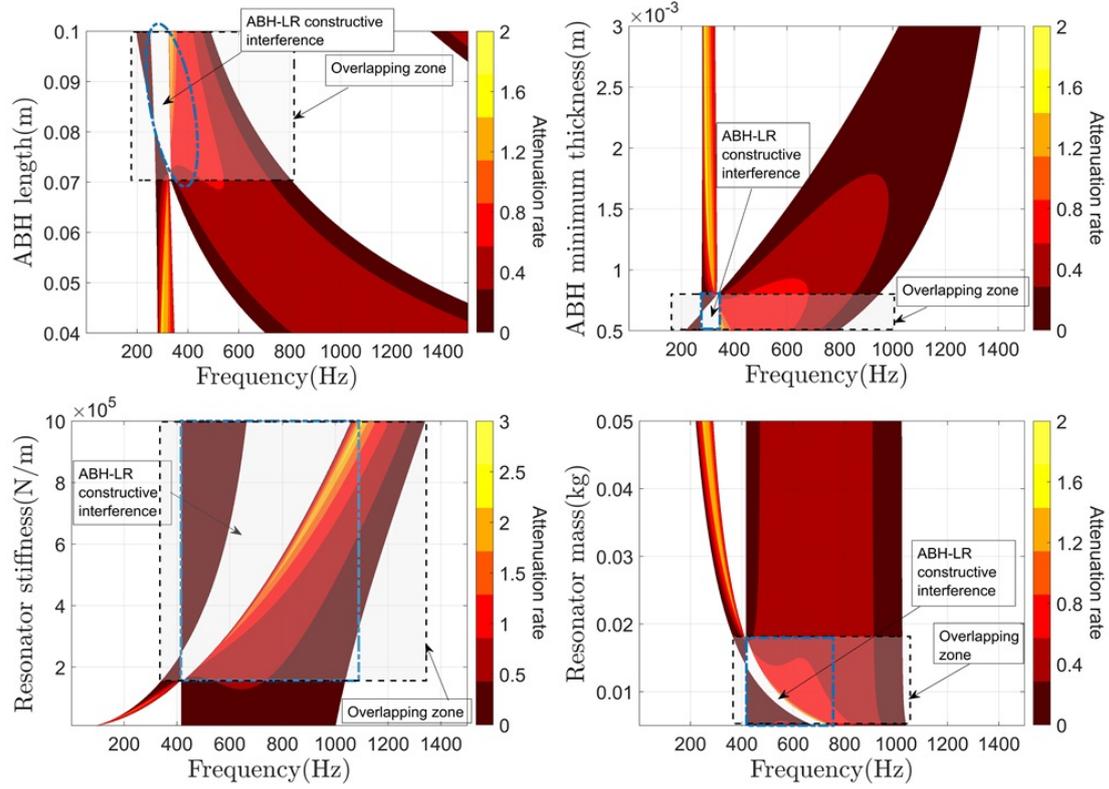

Figure 5.10: Attenuation profiles for the coupled ABHLR unit cell. Shaded region inside the rectangular area represents the overlapping zone. ABHLR constructive interference phenomenon is shown within the white region enclosed by the blue perimeter.

## 5.4 PARAMETRIC STUDIES OF ABHLR UNIT

Parametric studies of the dispersion results are shown in Fig. 5.10, termed as attenuation profile. The attenuation profiles are shown considering four parameters: parameters depending on the acoustic black hole (minimum thickness of beam at engraved ABH position $h_t$, length of engraved ABH $l_3 + l_4$), and parameters depending on local resonator (resonator mass $m_1$, resonator stiffness $k_1$). A single parameter is systematically adjusted to produce each attenuation profile while maintaining the other parameters constant.

- ABH attenuation band shifts toward the lower frequency range as the length of engraved ABH section ($l_3 + l_4$) increases, but the bandwidth reduces.

- ABH attenuation band shifts toward the lower frequency range and widens as the minimum thickness of the ABH section, denoted as $h_t$, decreases.

- As the stiffness of the local resonator ($k_1$) increases, the bandwidth of the pass-band widens, and the second attenuation band shifts towards the high-frequency region within the overlapping zone.

- LR attenuation band shifts towards the higher frequency region and results in a pass-band within the overlapping zone, as local resonator mass ($m_1$) decreases.

- Within the overlapping zone, the high attenuation rate of the second bandgap provides evidence for the transition of the LR effect into the high-frequency region. This transition is shown clearly in different parametric studies.

## 5.5 CONCLUSION

This research investigates the flexural wave propagation in a metamaterial beam with the coupled acoustic black hole and the local resonator to mitigate low-high frequency vibration. This is achieved by deriving the dynamic stiffness of coupled ABHLR unit cell, and its bandgap characteristics are analysed. Finally, the frequency domain response of a finite beam with ABHLR cells is analysed to validate the bandgap analysis of unit cell. The following conclusions are drawn:

1. For some parametric values, coupled ABHLR cell simultaneously produces low-frequency LR bandgap and high-frequency ABH bandgap. The LR bandgap creates almost 170% higher attenuation rate than the ABH bandgap but with a narrow region, at low frequencies. Meanwhile, the ABH produces a broader bandgap with lesser attenuation rate, at high frequencies.

2. For other parametric values, the coupled ABHLR cell shows a merged bandgap but with a pass-band appearing on the overlapping region. This pass-band appears due to constructive interference of ABH and LR. This phenomenon can be considered as a negative effect of coupled ABHLR bandgap formation.

3. The merged band shows a higher attenuation rate, compared to the uncoupled ABH or LR cell bandgaps. This higher attenuation rate appears due to the destructive interference of ABH and LR, and can be considered as a positive effect.

4. The coupled ABHLR cell exhibits a double peak in the merged attenuation band.

5. In the merged band gaps, the pass-band is highly sensitive to changes in resonator stiffness compared to other parameters of ABHLR cell. Careful design of the resonator stiffness is crucial for maximizing the coupling effects on bandgap formations.

6. Unit cell model of coupled ABHLR based on dynamic stiffness method is effective in capturing the essential phenomena and matches with the computationally expensive finite element simulations. Therefore, this coupled model can be used for design and optimization studies of ABHLR metamaterial.

This study shows that the coupling of ABH and LR metamaterial produces either a positive or negative effect on bandgap formations or vibration attenuation, depending on the ABHLR parameters. Therefore, the coupling has to be optimally designed to maximize the attenuation effects and avoid constructive interference.

# CHAPTER 6

# COUPLED ACOUSTIC BLACK HOLE & LOCAL RESONATOR: EXPERIMENTAL STUDY

## 6.1 INTRODUCTION

In the previous chapter, flexural wave propagation through coupled acoustic black hole and local resonators has been discussed. Numerical results indicate that the coupling of ABH and LR shows both positive and negative effects on band gap formations. Wang *et al.* (2023) conducted an experimental study on flexural vibration attenuation in phononic crystal beams and concluded that experimental results align with analytical findings when a sufficient number of unit cells are attached to the beam. Banerjee (2018) analyzed a bistable model experimentally and observed a discrepancy between computational and experimental results, attributing it to certain unavoidable and unmeasurable factors.

In this chapter, we investigate the experimental frequency response function (FRF) of a coupled ABHLR beam for the purpose of flexural vibration attenuation. To facilitate the comparison of the effectiveness of the coupled ABHLR beam, we also examine an uncoupled ABH beam. Both specimens are designed using COMSOL Multiphysics modelling and subsequently fabricated through 3D printing technology. This experimental study contributes to a thorough understanding of the dynamic behaviour and vibration attenuation capabilities of the ABHLR beam.

## 6.2 EXPERIMENTAL MODEL

To formulate the experimental model, a finite element simulation framework is established using COMSOL Multiphysics software. The downward and upward wedge of the ABH section are designed by varying the thickness along the x-direction in a parabolic fashion, similar to the approach discussed in Chapter 4. For the design of the local resonator in the coupled ABHLR beam, a cylindrical-type bending resonator is considered, as proposed by Mondal *et al.* (2023). Fig. 6.1 illustrates the 3D printed coupled ABHLR and uncoupled ABH beam specimens intended for experimental studies. Due to the size constraints of the 3D printing bed, the ABHLR or ABH beam with only three unit cells is constructed.

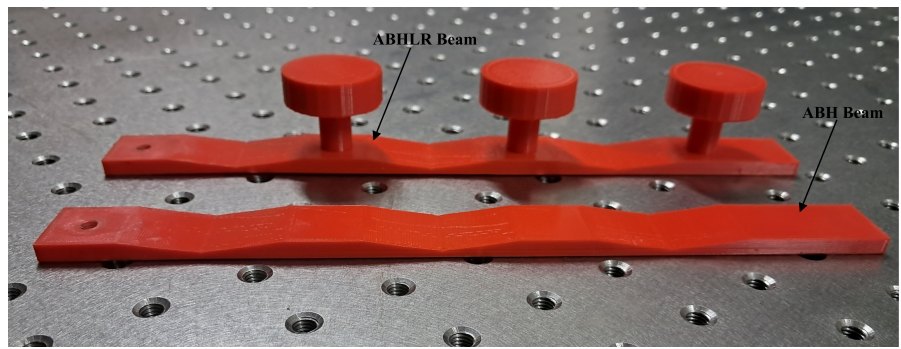

Figure 6.1: Coupled ABHLR and uncoupled ABH beam specimen for experimental studies.

Initially, numerical frequency response studies are conducted to evaluate the capability of these beams in attenuating flexural vibrations. Material properties are collected from the eSUN 3D printing filament properties data sheet obtained through wire testing. Since the wires are melted during the additive manufacturing process to build up the layers, the material properties of the printed specimen may vary slightly from its original material.

The geometrical parameters of the ABHLR or ABH beam include a length of 220 mm, a width of 20 mm, a maximum thickness of 5 mm, and a minimum thickness of 2 mm. These specimens are manufactured using PLA filament material through fused deposition modelling (FDM), a type of 3D printing technique. The material parameters of the ABHLR or ABH beam are presented in Table 6.1.

Table 6.1: Material Parameters of the ABH and ABHLR beam

| Properties | Unit |
|---|---|
| $\rho$ | $1250 kg/m^3$ |
| $E$ | 3600 MPa |
| $\mu$ | 0.33 |

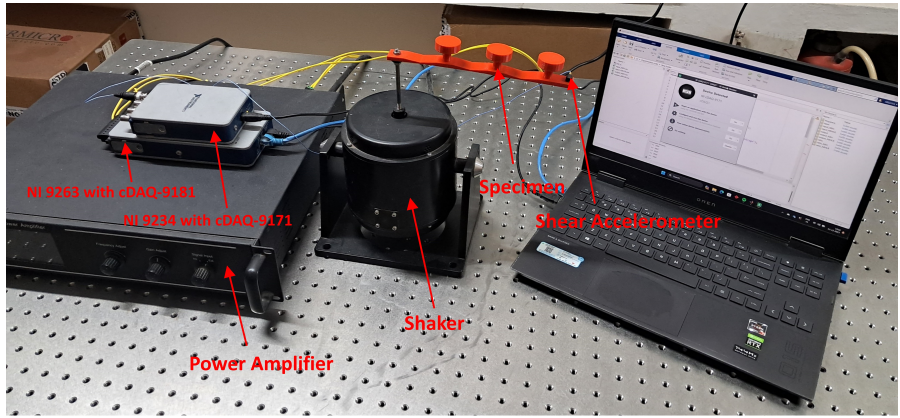

Figure 6.2: Experimental setup of FRF study for coupled ABHLR beam

## 6.3 EXPERIMENTAL SETUP

The experimental setup for the frequency response function study is depicted in Fig. 6.2. A MATLAB program was developed to generate a sine signal sweep within the frequency range of 1 Hz to 2000 Hz and to record sensor data for each frequency. As per the program specifications, signals were produced by the output DAQ module (NI 9263), connected to the laptop through the cDAQ-9181 chassis.

The ABHLR or ABH beam was subjected to excitation using the electromagnetic shaker (YMC MS-50), controlled by the generated signals and amplified by the power amplifier (YMC LA 100). The input excitation was applied to one end of the beam, and output data were acquired from the opposite side. Acceleration data for both input and output were measured using a shear accelerometer (PCB Piezotronics model: 352A24). These sensor data were collected by the input DAQ module (NI 9234), linked to the laptop via the cDAQ-9171 chassis.

Subsequently, the transmissibility of the beam was analyzed using the following formula:

$$T_{beam} = 20\log\frac{a_{out}}{a_{in}} \tag{6.1}$$

Where $T_{beam}$, $a_{out}$ and $a_{in}$ respectively denote the transmissibility and the acceleration amplitudes at the output and input points, respectively.

## 6.4 RESULTS & DISCUSSIONS

Transmissibility curves are plotted to check the vibration attenuation across ABH and ABHLR beams. Prior to specimen fabrication, transmissibility results are acquired through finite element simulation using COMSOL Multiphysics software. Subsequently, printed specimens are experimentally analyzed to determine the attenuation properties throughout beams.

### 6.4.1 Finite Element Simulations

The transmissibility curves obtained from COMSOL Multiphysics are illustrated in Fig. 6.3. A minimum transmissibility of approximately -32 dB is observed for the ABH beam, while for the ABHLR beam, it is approximately -71 dB. The ABHLR beam exhibits higher attenuation between the ranges of 240-720 Hz and 1615-2000 Hz compared to the ABH beam.

### 6.4.2 Experimental Results

Fig. 6.4 depicts the transmissibility curves of the printed ABHLR and ABH specimens obtained experimentally. The notable peak of 18 dB observed in the ABH specimen is reduced to 8 dB for the ABHLR specimen around a frequency of 300 Hz. A minimum transmissibility of approximately -13 dB is noted for the ABH beam, whereas for the ABHLR beam, it is approximately -16 dB. The first higher attenuation zone of the ABHLR specimen, as seen in simulation results, is not as clearly visible in the experimental findings. However, the ABHLR specimen exhibits a prominent bandgap zone in the 930-2000 Hz range, surpassing that of the ABH specimen. The experimental results also affirm the superiority of the ABHLR specimen over the ABH specimen.

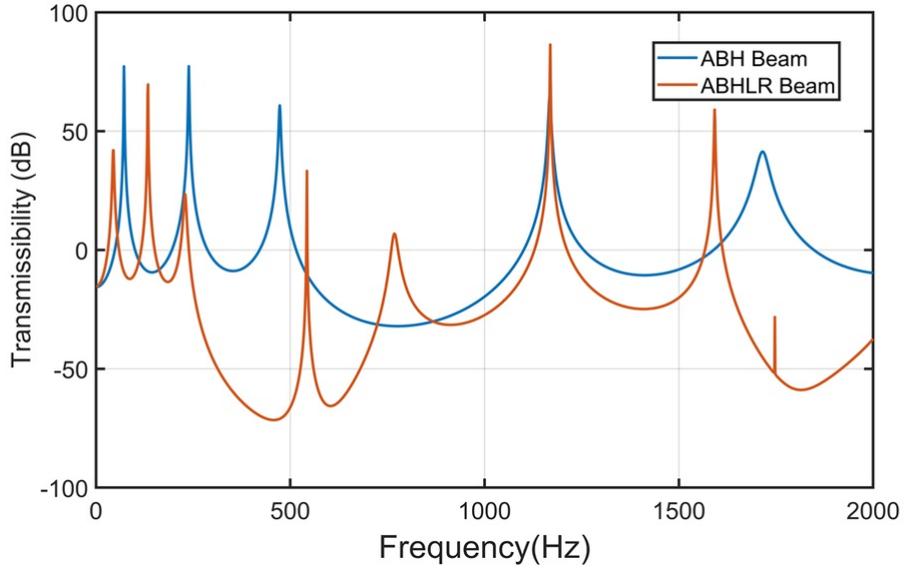

Figure 6.3: COMSOL simulations results of FRF study

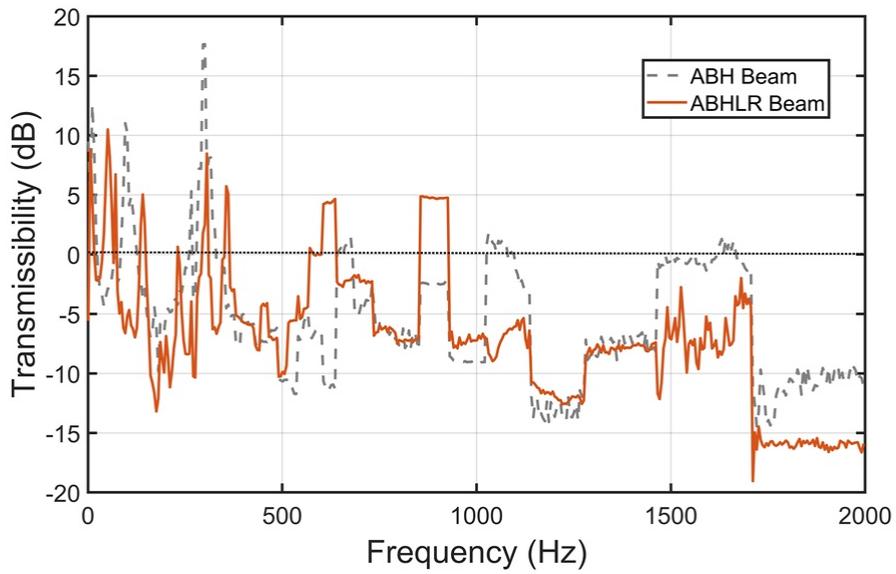

Figure 6.4: Experimental results of FRF study

### 6.4.3 Comparison with COMSOL Results

It is observed that computational and experimental results do not precisely align. The potential reasons are outlined as follows:

- The material properties are obtained from the filament material properties data sheet. However, it is essential to acknowledge that material properties may vary due to manufacturing defects.

- The numerical model assumes the material property as elastic and linear, but in reality, the properties of the 3D printed polymer-based samples may not be entirely elastic.

- The damping of the 3D printed sample is relatively high and unknown. The numerical model does not account for any assumed damping.

- For very low-amplitude vibrations, the displacement amplitude is very small. Thus, the presence of environmental noise can have a significant impact on them.

- For high-amplitude vibrations at low frequencies, nonlinearity can be introduced into the systems. However, for the numerical study, it is not considered.

### 6.5 CONCLUSIONS

This chapter explores the experimental study of flexural vibration attenuation of a metamaterial beam with the coupled acoustic black hole and local resonator. The coupled ABHLR and uncoupled ABH beam specimens are manufactured using 3D printing technology, and experimental frequency response results are obtained. The following conclusions are drawn:

1. In comparison to the uncoupled ABH beam, the ABHLR beam exhibits a better capacity for vibration attenuation across a wider frequency range.

2. Despite encountering several challenges in comparison with simulation results, experimental findings highlight the advantage of the ABHLR cell.

3. The coupling of a metamaterial beam with ABH and LR can be employed for wide-frequency vibration attenuation.

# CHAPTER 7

# MULTISTABLE OSCILLATOR: NUMERICAL STUDY

## 7.1 INTRODUCTION

Nonlinear metamaterials have emerged as a promising area of research in structural dynamics, specifically for structural vibration suppression. Unlike traditional linear resonator metamaterials, nonlinear oscillator units can exhibit intriguing and controllable nonlinear responses, enabling a wide range of attenuation band regions. Nonlinear resonance can increase bandgaps by using properties such as sub and super-harmonic resonances, period multiplication, and chaotic response. Multistability is a nonlinear resonance phenomenon that is influenced by disturbances, initial conditions, system characteristics, and other factors. According to Banerjee *et al.* (2016), the effects of cubic duffing type non-linearity on 1D elastodynamic metamaterials revealed that a bistable nonlinear system has a broader attenuation bandwidth than a monostable system. Xia *et al.* (2019) observed that a base-excited cantilever beam with magneto-elastic bistable unit cells has a much wider attenuation bandwidth due to nonlinear interwell oscillations than the linear locally resonant bandgap.

This chapter shows a comparative analysis of nonlinear multistable oscillators in a 1D metamaterial chain to understand the insights between the oscillator's potential well and the metamaterial chain's attenuation band. Bloch's theorem cannot be directly used to solve nonlinear metamaterial because of the amplitude dependency of each consecutive oscillator. A numerical time-domain analysis is applied using the Range-Kutta method (MATLAB ODE45) for solving finite nonlinear metamaterial chains, as discussed by Xia *et al.* (2019), and frequency response (up and down frequency sweep) results at each frequency are collected using root mean square values of steady-state displacements. Finally, a correlation is observed by the comparative analysis of different kinds of nonlinear oscillators.

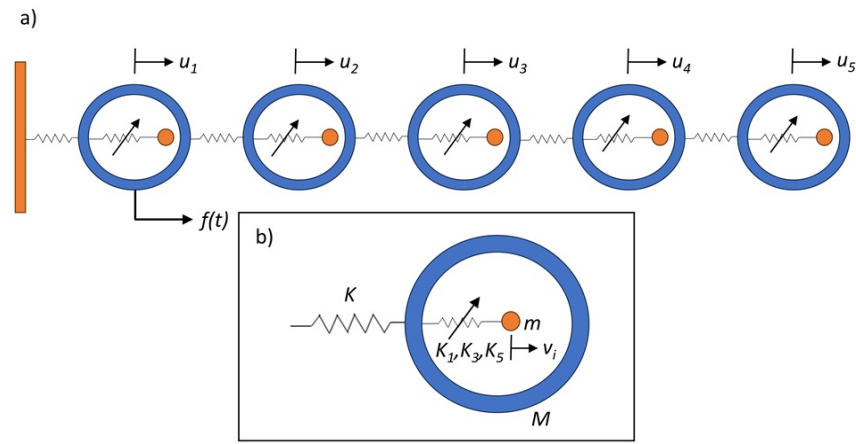

Figure 7.1: a) Finite metamaterial chain with multistable oscillators b) Single unit cell of metamaterial chain. All the parameters of the system are shown in the figure.

## 7.2 METHODOLOGY

In this study, a linear spring-mass metamaterial chain with multistable nonlinear oscillators is considered for a basic qualitative concept demonstration. Three types of nonlinear oscillators are presented depending on the stability points to compare the effect of attenuation bands with the increase of stable equilibrium positions:

1. Monostable oscillator

2. Bistable oscillator

3. Tristable oscillator

Fig. 7.1 represents a finite metamaterial chain with five multistable oscillators. We have considered 10 degrees of freedom system with five main masses M and five multistable oscillators with mass m, as discussed by Xia *et al.* (2019). The stiffness of the main spring chain is denoted as K, while the stiffness of the multistable oscillators can be characterized by linear stiffness $K_1$, cubic nonlinear stiffness $K_3$, and fifth-order nonlinear stiffness $K_5$.

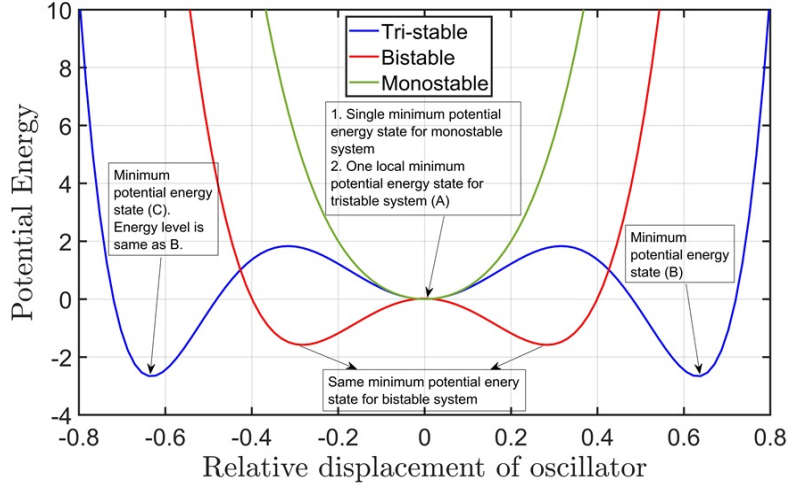

Figure 7.2: Potential energy wells for nonlinear multistable oscillators

The governing equations of motion for the $i^{th}$ main mass and the $i^{th}$ oscillator mass are:

$$M\ddot{u}_i + C_1(\dot{u}_i - \dot{u}_{i-1}) + C_1(\dot{u}_i - \dot{u}_{i+1}) + K(u_i - u_{i-1}) + K(u_i - u_{i+1}) \\ + C_2(\dot{u}_i - \dot{v}_i) + K_1(u_i - v_i) + K_3(u_i - v_i)^3 + K_5(u_i - v_i)^5 = f(t)\delta_{i1} \quad (7.1)$$

$$m\ddot{v}_i + C_2(\dot{v}_i - \dot{u}_i) + K_1(v_i - u_i) + K_3(v_i - u_i)^3 + K_5(v_i - u_i)^5 = 0 \quad (7.2)$$

Where $u_i$ is the absolute displacement of the $i^{th}$ main mass, and $v_i$ is the displacement of the $i^{th}$ oscillator mass. $C_1$ and $C_2$ represent the damping coefficients of the main spring-mass chain and the multistable oscillators, respectively. The leftmost side of the metamaterial chain remains fixed, and a time-varying force $f(t) = F\sin(\omega t)$ is applied at the first metamaterial unit cell. Here, $\delta_{i1}$ represents the Kronecker delta function. Table 7.1 shows the values of stiffness and points of stability for the monostable, bistable, and multistable resonators.

Fig. 7.2 shows the potential energy of the three types of nonlinear oscillators. The bandwidth of the nonlinear system is sensitive to the initial displacement of the oscillator attachments. Therefore, calculating potential energy plays a crucial role in determining the stable initial displacement. The restoring force $F(x)$ and the potential energy $U(x)$ can be calculated using the following formulas:

$$F(x) = K_1 x + K_3 x^3 + K_5 x^5 \quad (7.3)$$

$$U(x) = \frac{K_1 x^2}{2} + \frac{K_3 x^4}{4} + \frac{K_5 x^6}{6} + C \quad (7.4)$$

Table 7.1: Different parameters for nonlinear multistable oscillators Banerjee (2018).

| Coefficients | Monostable | Bistable | Tristable |
|---|---|---|---|
| $K_1$ | positive | negative | positive |
| $K_3$ | positive | positive | negative |
| $K_5$ | 0 | 0 | positive |
| Point of stability | 0 | $\pm\sqrt{\frac{K_1}{K_3}}$ | $0, \pm\sqrt{\frac{K_3+\sqrt{K_3^2-4K_5K_1}}{2K_5}}$ |
| C | 0 | $\sqrt{\frac{K_3}{16K_1}}$ | 0 |

Where $C$ is the coefficient. It has been shown in Table 7.1. This coefficient can vary depending on the stability of the oscillators. A numerical solution method is applied using the MATLAB ODE45 throughout the frequency domain of interest, and the root mean square (RMS) values of steady-state displacement amplitudes are collected for each up and down frequency sweep, as discussed by Xia *et al.* (2019). The state-space form of a 10-DOF meta-structure has been shown in Appendix A. The size of the state space matrix in this system with 5 unit cells is ten by ten. Similarly, the increase in the number of unit cells (n) will substantially increase the size of the state space matrix (4n) and making the analysis complicated.

The steady-state amplitudes are reached near 4000 cycles of oscillations, and RMS values are taken from these regions. For this study, the first-frequency excitation is applied from the initial stable positions of the multistable oscillators, and the initial position for the next-frequency excitation is taken from the immediate last position of the multistable oscillators after completing 4000 cycles. This is considered as frequency sweep for this study. The initial stable positions of multistable oscillators for different types of oscillations are given in Table 7.2. The attenuation bandwidth of the metamaterial chain can be computed based on the transmittance $\tau$, which is the ratio of the displacement amplitude of the rightmost unit cell $u_5$ and the displacement amplitude of the leftmost unit cell $u_1$. This can be represented by

$$\tau = 20 \log \left( \frac{u_5}{u_1} \right) \tag{7.5}$$

Table 7.2: Initial position of nonlinear multistable oscillators.

| Case No. | Oscillation Type | Monostable | Bistable | Tristable |
|---|---|---|---|---|
| Case 1 | Intrawell | 0 | $\sqrt{\frac{K_1}{K_3}}$ | 0 |
| Case 2 | Intrawell | - | $-\sqrt{\frac{K_1}{K_3}}$ | $\sqrt{\frac{K_3+\sqrt{K_3^2-4K_5 K_1}}{2K_5}}$ |
| Case 3 | Interwell | - | $\sqrt{\frac{K_1}{K_3}}$ | 0 |

This numerical case study considers a main mass of $M = 0.1$ kg, a spring stiffness of the metamaterial chain $K = 100$ N/m, a oscillator mass $m = 0.1$ kg, and damping coefficients $C_1 = C_2 = 0.2$ Ns/m. Simulations are carried out for both up and down frequency sweeps throughout the frequency range between 1 Hz and 20 Hz in order to evaluate the frequency response function of the nonlinear system. This study considers two forcing amplitudes: one is $F = 4 \times 10^{-1}$ N for intrawell oscillations, and the other one is $F = 4 \times 10^1$ N for interwell oscillations. The stiffness values of nonlinear oscillators are shown below:

- Monostable system: $K_1 = 80$ N/m, $K_2 = 1000$ N/m
- Bistable system: $K_1 = -80$ N/m, $K_2 = 1000$ N/m

- Tristable system: $K_1 = 80$ N/m, $K_2 = -1000$ N/m, $K_3 = 2000$ N/m

## 7.3 RESULTS AND DISCUSSIONS

Nonlinear oscillators are systematically compared from a stability perspective to elucidate the insights into multistable phenomena for bandwidth enhancement. Multistable oscillators generate multiple potential wells, the characteristics of which hinge on the number of stability points exhibited by the oscillators. These potential wells can be categorized into two types based on the minimum potential energy and the steepness of the curve. Identical potential wells share the same type of minimum potential energy and curve steepness, while nonidentical wells exhibit differing types of potential energy and curve steepness. In light of amplitude dependency, a nonlinear dynamic system manifests two distinct types of oscillations for multistable oscillators.

- Intrawell oscillations arise from a low level of excitations, wherein the energy imparted by the excitation is insufficient to overcome the separatrix energy of the potential well.

- Interwell oscillations manifest in response to high-level excitations, where the attained energies surmount the separatrix energy of the potential well. This type of oscillation can induce chaotic motion as it traverses through all the potential wells of the oscillators.

Monostable, bistable, and tristable oscillators are individually examined, and correlations with potential wells are discerned.

### 7.3.1 Monostable oscillator

The monostable oscillator features singular potential energy well, constraining excitations to intrawell oscillations. The potential well of the monostable oscillator is depicted in Fig. 7.2, while Fig. 7.3 illustrates the transmissibility plot of the finite metamaterial chain incorporating monostable oscillators. An attenuation band, exhibiting an almost 2.2 Hz bandwidth, is observed for both low and high levels of excitations, attributed to the presence of a single stability point. As excitation levels increase, the system's transmissibility rises even within the bandgap zone, unlike a linear metamaterial chain. During up-sweeps and down-sweeps of frequency, no significant difference is noted in the transmissibility plot for the monostable system.

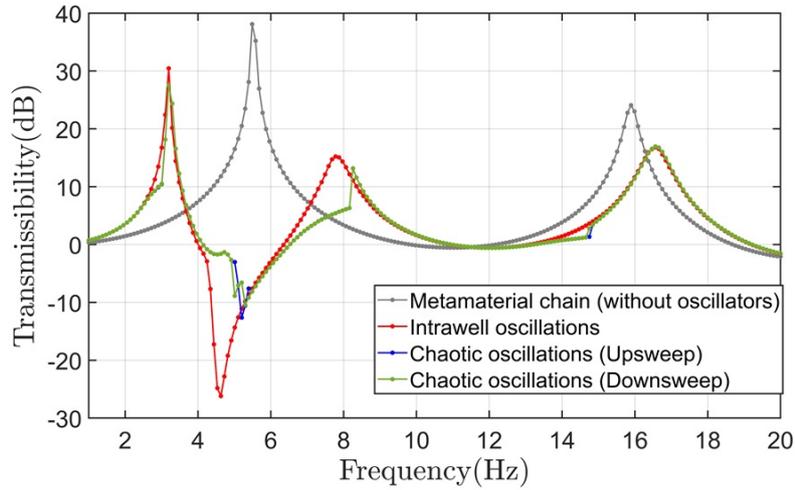

Figure 7.3: Transmissibility plot of the metamaterial chain with monostable oscillators.

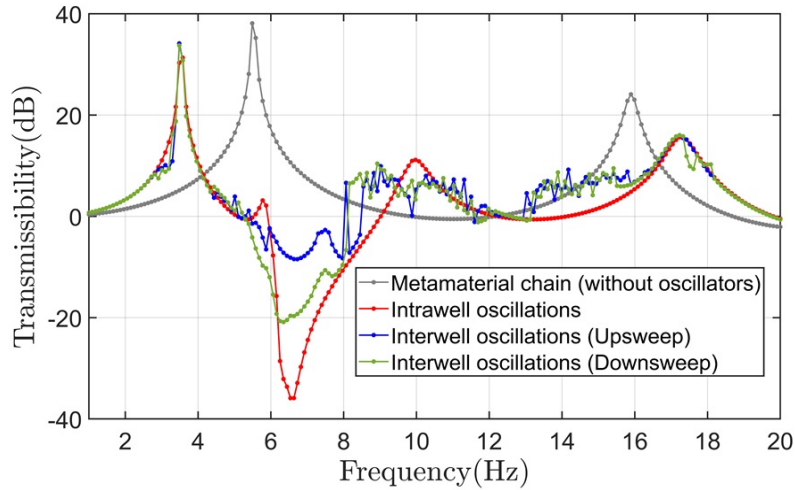

Figure 7.4: Transmissibility plot of the metamaterial chain with bistable oscillators.

### 7.3.2 Bistable oscillator

The bistable oscillator exhibits two stability points or potential wells, both of which are identical. Fig. 7.2 illustrates the potential well for a bistable oscillator, and Fig. 7.4 presents the transmissibility plot of the metamaterial chain featuring bistable oscillators. The energy levels and curve steepness within these potential wells for the bistable oscillator are identical. Consequently, an identical attenuation band is observed at any initial stable position in the case of intrawell oscillations. The transition between the two potential wells occurs under high excitation, leading to interwell oscillations. However, even for interwell oscillations, a single attenuation band is observed due to the identical potential wells.

The figure depicts an attenuation bandwidth of 2.5 Hz attributed to intrawell oscillations. The frequency response of the bistable system demonstrates sensitivity to both up-sweep and down-sweep frequencies during interwell oscillations. Interwell oscillations produce nearly the same attenuation band as intrawell oscillations, with minor changes for the bistable oscillator. However, the transmissibility of the metamaterial chain increases with high excitations, consistent with previous observations.

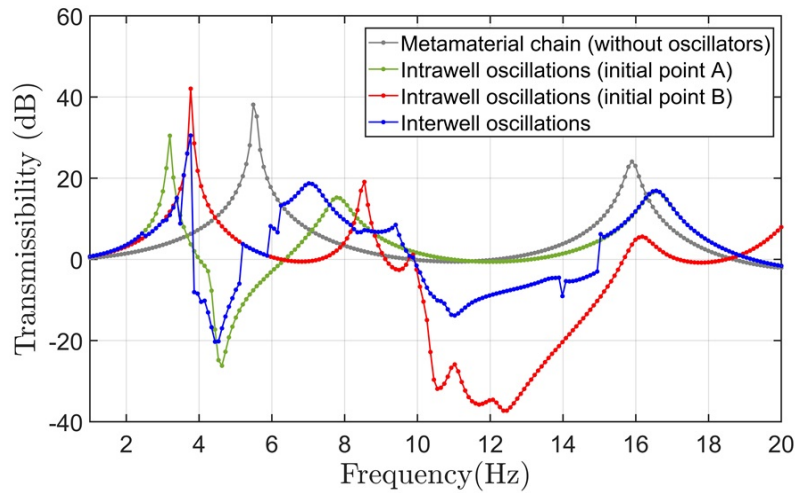

Figure 7.5: Transmissibility plot of the metamaterial chain with tristable oscillators. This figure presents an FRF plot for the up-frequency sweep.

### 7.3.3 Tristable oscillator

The nonlinear tristable oscillator features three stability points or potential wells, with two of them exhibiting nonidentical potential energy wells. The potential wells for the tristable oscillator are illustrated in Fig. 7.2. Three stable minimum potential energy levels are identified as Position A, Position B, and Position C. Positions B and C give rise to identical potential energy wells for the resonator, owing to their equivalent minimum potential energy levels and curve steepness.

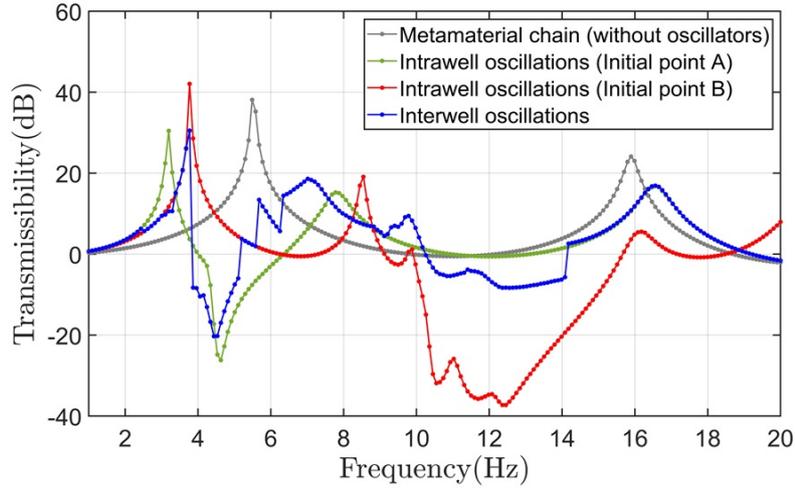

Figure 7.6: Transmissibility plot of the metamaterial chain with tristable oscillators. This figure presents an FRF plot for the down-frequency sweep.

Fig. 7.5 and Fig. 7.6 depict the transmissibility plots for up-sweep and down-sweep frequencies of the tristable oscillator. An attenuation band between 4 Hz and 6.2 Hz is evident due to intrawell oscillations with the initial stable position at A. Another distinct attenuation band, ranging from 10 Hz to 15.5 Hz, is observed due to intrawell oscillations with the initial stable position at B or C. Simultaneously, two bandgap regions between 3.8 to 5.2 Hz and 10 to 15 Hz emerge due to interwell oscillations of the tristable oscillator.

The attenuation band of the metamaterial chain is found to depend on the potential well of the initial stability point in the case of intrawell oscillations. Two distinct attenuation bands arise due to two nonidentical potential wells in the system for interwell oscillations. Furthermore, it is observed that interwell oscillations are sensitive to the frequency sweeping direction for the tristable oscillator.

## 7.4 CONCLUSIONS

A comprehensive investigation of a metamaterial chain incorporating nonlinear multistable oscillators is conducted to improve the bandwidth region. Monostable, bistable, and tristable oscillators are specifically analyzed to discern the correlation between the potential well characteristics of multistable resonators and the attenuation bandwidth of metamaterial structures. Employing a time-domain numerical analysis, the impact of nonlinear multistable oscillators on a 10-DOF metamaterial chain is systematically explored. The key conclusions derived from this study are as follows:

1. Intrawell oscillations of a multistable oscillator can induce a singular attenuation band contingent upon the initial stability point. Conversely, interwell oscillations generate multiple attenuation bands by traversing through distinct stability points.

2. Potential wells characterized by the same minimum potential energy and curve steepness are termed identical potential wells. Attenuation bands arising from two identical potential wells are identical, while two nonidentical potential wells give rise to distinct attenuation bands.

3. Attenuation bands resulting from interwell oscillations are contingent upon the nonidentical potential wells of the oscillator. The number of bandgap regions aligns with the number of nonidentical potential wells.

4. Nonlinear multistable oscillators exhibit the capability to open multiple attenuation bands, thereby enhancing vibration suppression, particularly in mechanical and aerospace applications.

This chapter represents the initial study of multistable oscillators in metamaterial structure to understand the multistable behaviours. In future, these multistable oscillators can be coupled with acoustic black holes in metamaterial beam for flexural vibration attenuation.

# CHAPTER 8

# CONCLUSIONS & FUTURE DIRECTIONS

In this chapter, we summarize the overall conclusions of the thesis and outline future directions. Detailed conclusions for each study have been discussed in their respective chapters.

## 8.1 CONCLUSIONS

Throughout this thesis, we have studied the dynamics of coupled metamaterials to achieve vibration attenuation across multiple frequencies. In the first part, a metamaterial beam with the coupled acoustic black hole and local resonator is analysed to enhance flexural vibration attenuation in low-high frequency regions. These coupled metamaterials show both positive and negative effects in bandgap formations. As a negative effect, for some parametric values, the ABH and LR bandgaps merge but also produce a passband on the overlapping region. So, the coupling has to be optimally designed to maximize the attenuation effects and to avoid passband.

In the second part, we explored a metamaterial chain featuring multistable oscillators for the purpose of longitudinal vibration attenuation. Our findings uncovered a unique correlation between the attenuation band of the metamaterial chain and the potential well of a nonlinear multistable oscillator. For interwell oscillations, the number of attenuation bands depends on the nonidentical potential well of the multistable oscillator. These types of metamaterials hold great promise for applications in mechanical and aerospace engineering, offering the capability for effective vibration attenuation across multiple frequencies. This research contributes valuable insights into the utilization of coupled metamaterials for enhancing performance in various engineering applications.

## 8.2 FUTURE DIRECTIONS

The aforementioned conclusion elucidates that coupled metamaterials can effectively serve the purpose of attenuating vibrations across multiple frequencies. As a future work, the optimization of coupled ABHLR metamaterial holds the potential to enhance its attenuation properties while mitigating any adverse effects. Further, nonlinear multistable oscillators can be coupled with acoustic black holes in the metamaterial beam for flexural vibration attenuation. It has the capability to open up new opportunities in the formation of attenuation bands. Considering all these prospective aspects, this study lays a foundation for exploring new opportunities in coupled metamaterial research.

# APPENDIX A

# APPENDIX

## A.1 EXACT SOLUTION OF ABH SUB-UNIT

The governing equation of Euler- Bernoulli's beam for a particular frequency $\omega$ can be written as

$$\frac{d^2}{dx^2}\left(EI(x)\frac{d^2W}{dx^2}\right) - \rho A(x)\omega^2 W = 0 \tag{A.1}$$

For the ABH: Downwards wedge sub-unit, A.1 can be written as after putting $I(x)$ and $A(x)$:

$$\frac{Ebh_o^3}{12}\frac{d^2}{dx^2}\left((1-\frac{x}{x_o})^6\frac{d^2W}{dx^2}\right) - \rho bh_o(1-\frac{x}{x_o})^2\omega^2 W = 0$$
$$\Rightarrow (1-\frac{x}{x_o})^4\frac{d^4W}{dx^4} - \frac{12}{x_o}(1-\frac{x}{x_o})^3\frac{d^3W}{dx^3} + \frac{30}{x_o^2}(1-\frac{x}{x_o})^2\frac{d^2W}{dx^2} - \frac{12\rho\omega^2}{Eh_o^2}W = 0 \tag{A.2}$$

Assuming the solution type $W = (1-\frac{x}{x_o})^m$ in A.2, we can write

$$m(m-1)(m-2)(m-4)\frac{1}{x_o^4} + \frac{12}{x_o^4}m(m-1)(m-2) + \frac{30}{x_o^4}m(m-1) - \frac{12\rho\omega^2}{Eh_o^2} = 0 \tag{A.3}$$

By solving the above equation, we can write the solution as:

$$W(x) = a_1\left(1-\frac{x}{x_0}\right)^{m_1} + a_2\left(1-\frac{x}{x_0}\right)^{m_2} + a_3\left(1-\frac{x}{x_0}\right)^{m_3} + a_4\left(1-\frac{x}{x_0}\right)^{m_4} \tag{A.4}$$

where $m_1, m_2, m_3$ and $m_4$ are the function of the particular frequency $\omega$. These can be represented by

$$\begin{aligned} m_1 &= -\frac{3}{2} - \frac{1}{2}\sqrt{17 - 4\sqrt{4+\lambda}} & m_2 &= -\frac{3}{2} - \frac{1}{2}\sqrt{17 + 4\sqrt{4+\lambda}} \\ m_3 &= -\frac{3}{2} + \frac{1}{2}\sqrt{17 - 4\sqrt{4+\lambda}} & m_4 &= -\frac{3}{2} + \frac{1}{2}\sqrt{17 + 4\sqrt{4+\lambda}} \end{aligned} \tag{A.5}$$

where $\lambda = \frac{12x_0^4\rho\omega^2}{Eh_0^2}$. A.4 represents the exact solution of ABH: Downwards wedge sub-unit. The exact solution for ABH: Upwards wedge sub-unit can be derived in the same way as before.

## A.2  STATE-SPACE OF MULTISTABLE OSCILLATOR

To solve the metamaterial chain with multistable oscillators, we have written the state-space form of a 10-DOF system in MATLAB program. The mentioned state-space form is shown below:

```
%du represents the state-space representation of the 10-
    DOF system
% u(4n-3) represents the displacement of n^th main mass
    system.
% u(4n-2) represents the velocity of n^th main mass
    system.
% u(4n-1) represents the displacement of n^th oscillator
    mass system.
% u(4n) represents the velocity of n^th oscillator mass
    system.
% P is the amplitude of excitation
% E1 and E2 are the damping coefficients of the main and
    oscillator mass systems, respectively.
% k is the stiffness of the linear main spring-mass chain
    .
% ka, ka3, ka5 are the first, third and fifth-order
    stiffness values according to the type of stability.

du = [u(2);
```

| | |
|---|---|
| 12 | `(P*sin(w*t)-E1*(u(2)-u(6))-E1*u(2)-k*(u(1)-u(5))-k3*(u(1)-u(5))^3-k*u(1)-k3*u(1)^3-E2*(u(2)-u(4))-ka*(u(1)-u(3))-ka3*(u(1)-u(3))^3-ka5*(u(1)-u(3))^5)/M;` |
| 13 | `u(4);` |
| 14 | `(-E2*(u(4)-u(2))-ka*(u(3)-u(1))-ka3*(u(3)-u(1))^3-ka5*(u(3)-u(1))^5)/m;` |
| 15 | `u(6);` |
| 16 | `(-E1*(u(6)-u(2))-E1*(u(6)-u(10))-k*(u(5)-u(1))-k3*(u(5)-u(1))^3-k*(u(5)-u(9))-k3*(u(5)-u(9))^3-E2*(u(6)-u(8))-ka*(u(5)-u(7))-ka3*(u(5)-u(7))^3-ka5*(u(5)-u(7))^5)/M;` |
| 17 | `u(8);` |
| 18 | `(-E2*(u(8)-u(6))-ka*(u(7)-u(5))-ka3*(u(7)-u(5))^3-ka5*(u(7)-u(5))^5)/m;` |
| 19 | `u(10);` |
| 20 | `(-E1*(u(10)-u(6))-E1*(u(10)-u(14))-k*(u(9)-u(5))-k3*(u(9)-u(5))^3-k*(u(9)-u(13))-k3*(u(9)-u(13))^3-E2*(u(10)-u(12))-ka*(u(9)-u(11))-ka3*(u(9)-u(11))^3-ka5*(u(9)-u(11))^5)/M;` |
| 21 | `u(12);` |
| 22 | `(-E2*(u(12)-u(10))-ka*(u(11)-u(9))-ka3*(u(11)-u(9))^3-ka5*(u(11)-u(9))^5)/m;` |
| 23 | `u(14);` |
| 24 | `(-E1*(u(14)-u(10))-E1*(u(14)-u(18))-k*(u(13)-u(9))-k3*(u(13)-u(9))^3-k*(u(13)-u(17))-k3*(u(13)-u(17))^3-E2*(u(14)-u(16))-ka*(u(13)-u(15))-ka3*(u(13)-u(15))^3-ka5*(u(13)-u(15))^5)/M;` |

```
25        u(16);
26        (-E2*(u(16)-u(14))-ka*(u(15)-u(13))-ka3*(u(15)-u
              (13))^3-ka5*(u(15)-u(13))^5)/m;
27        u(18);
28        (-E1*(u(18)-u(14))-k*(u(17)-u(13))-k3*(u(17)-u(13))
              ^3-E2*(u(18)-u(20))-ka*(u(17)-u(19))-ka3*(u(17)-u
              (19))^3-ka5*(u(17)-u(19))^5)/M;
29        u(20);
30        (-E2*(u(20)-u(18))-ka*(u(19)-u(17))-ka3*(u(19)-u
              (17))^3-ka5*(u(19)-u(17))^5)/m;
31        ]
```

# CURRICULUM VITAE

**NAME**                Arghya Mondal

**DATE OF BIRTH**       20 September 1998

**EDUCATION QUALIFICATIONS**

| | | |
|---|---|---|
| **2020** | **Bachelor of Technology** | |
| | Institution | Kalyani Government Engineering College |
| | Specialization | Mechanical Engineering |
| | | |
| **2024** | **Master of Science** | |
| | Institution | Indian Institute of Technology Madras |
| | Specialization | Aerospace Engineering |
| | Registration Date | 2 August, 2021 |

# GENERAL TEST COMMITTEE

**Chairperson**         Dr. H S N Murthy
                               Aerospace Engineering
                               Indian Institute of Technology Madras

**Guide**         Dr. Senthil Murugan
                               Aerospace Engineering
                               Indian Institute of Technology Madras

**Member(s)**         Dr. David Kumar
                               Aerospace Engineering
                               Indian Institute of Technology Madras

                               Dr. Ganesh Tamadapu
                               Applied Mechanics
                               Indian Institute of Technology Madras